\documentclass{article}

\usepackage{microtype}
\usepackage{graphicx}
\usepackage{subcaption}
\usepackage{booktabs} 

\usepackage{placeins}
\usepackage{tcolorbox}
\usepackage{mathrsfs}

\usepackage{hyperref}
\usepackage{longtable}

\usepackage[preprint]{icml2026}

\usepackage{amsmath}
\usepackage{amssymb}
\usepackage{mathtools}
\usepackage{amsthm}

\newcommand{\cref}[1]{\ref{#1}}
\newcommand{\Cref}[1]{\ref{#1}}

\theoremstyle{plain}

\theoremstyle{definition}

\theoremstyle{remark}

\newcommand{\todo}[1]{}
\newcommand{\missingfigure}[1]{}

\icmltitlerunning{Theorem Search}

\begin{document}

\twocolumn[
\icmltitle{Semantic Search over 9 Million Mathematical Theorems}



\icmlsetsymbol{equal}{*}

\begin{icmlauthorlist}
\icmlauthor{Luke Alexander}{equal,yyy,mathdep}
\icmlauthor{Eric Leonen}{equal,yyy,eric}
\icmlauthor{Sophie Szeto}{equal,yyy,mathdep,sophie}
\icmlauthor{Artemii Remizov}{yyy,artemii}
\icmlauthor{Ignacio Tejeda}{yyy,mathdep}
\icmlauthor{Jarod Alper}{yyy,mathdep}
\icmlauthor{Giovanni Inchiostro}{yyy,mathdep}
\icmlauthor{Vasily Ilin}{yyy,mathdep}
\end{icmlauthorlist}

\icmlaffiliation{yyy}{Math AI Lab, University of Washington, Seattle, United States}
\icmlaffiliation{mathdep}{Department of Mathematics, University of Washington, Seattle, United States}
\icmlaffiliation{sophie}{Paul G. Allen School of Computer Science and Engineering, University of Washington, Seattle, United States}
\icmlaffiliation{artemii}{Lake Washington High School, Kirkland, United States}
\icmlaffiliation{eric}{Department of Applied and Computational Mathematical Sciences, University of Washington, Seattle, United States}

\icmlcorrespondingauthor{Vasily Ilin}{vilin@uw.edu}

\icmlkeywords{semantic search, mathematical information retrieval, theorem retrieval, text embeddings, large language models}

\vskip 0.3in
]



\printAffiliationsAndNotice{\icmlEqualContribution} 

\begin{abstract}
Searching for mathematical results remains difficult: most existing tools retrieve entire papers, while mathematicians and theorem-proving agents often seek a specific theorem, lemma, or proposition that answers a query. While semantic search has seen rapid progress, its behavior on large, highly technical corpora such as research-level mathematical theorems remains poorly understood. In this work, we introduce and study semantic theorem retrieval at scale over a unified corpus of $9.2$ million theorem statements extracted from arXiv and seven other sources, representing the largest publicly available corpus of human-authored, research-level theorems. We represent each theorem with a short natural-language description as a retrieval representation and systematically analyze how representation context, language model choice, embedding model, and prompting strategy affect retrieval quality. On a curated evaluation set of theorem-search queries written by professional mathematicians, our approach substantially improves both theorem-level and paper-level retrieval compared to existing baselines, demonstrating that semantic theorem search is feasible and effective at web scale. The project page, search tool, dataset, REST API, and MCP server are available at \href{https://theoremsearch.com}{theoremsearch.com}.
\end{abstract}

\section{Introduction}

Mathematical knowledge is organized around discrete results: theorems, lemmas, propositions, and corollaries. These statements serve as the fundamental units of reasoning for both human mathematicians and automated proof systems \cite{polu2020gptf,yang2023leandojo,wu2022autoformalization}. A researcher proving a new result must first determine whether the statement already exists in the literature, and similarly, an AI agent generating formal proofs benefits from retrieving relevant lemmas to guide its search. Yet most existing tools -- Google Scholar, arXiv, and even modern LLMs with web access -- operate at the level of entire documents, forcing users to manually scan papers when they seek a specific statement.

This gap is increasingly significant. arXiv hosts over 2.4 million papers, including more than 690,000 in mathematics \cite{Ginsparg1994ArXiv}. A study of over 14,000 withdrawn arXiv preprints found that 2.5\% were retracted because the authors' results already appeared in prior literature \cite{withdarxiv2024}. For example, \citet{popescu2007stein}, \citet{zhang2012noncongruent}, and \citet{shahryari2020surjunctive} were withdrawn after discovering their main results had been previously established. AI systems face the same problem: the Erd\H{o}s Problems Project documented cases where AI tools ``solved'' open problems that had been established decades earlier \cite{erdos1977bases,wirsing1961approximation,klarner1966representations} -- underscoring the need for theorem-level search.


In this work, we construct a corpus of over 9 million theorem statements from arXiv, the Stacks Project, ProofWiki, and five other sources, and study semantic retrieval at scale. We represent each theorem using a natural-language ``slogan'' generated by an LLM, then embed slogans and queries into a shared semantic space. Our main contributions are:
\begin{enumerate}
    \item \textbf{A large-scale theorem corpus.} We release over 9 million theorem statements with rich metadata, the \textit{largest} collection of informal mathematical theorems to date.
    \item \textbf{A systematic study of representation choices.} We analyze how the context, LLM choice, embedding model, and prompting strategy affect retrieval. Notably, we find that embedding theorems via natural-language slogans significantly outperforms embedding their raw \LaTeX{} formulations.
    \item \textbf{State-of-the-art retrieval.} On 111 queries from professional mathematicians, we achieve 45.0\% Hit@20 at the theorem level, outperforming ChatGPT 5.2 with search (19.8\%) and Gemini 3 Pro (27.0\%). For paper-level retrieval, we achieve 56.8\% Hit@20 compared to 37.8\% for Google Search.
\end{enumerate}

Our results demonstrate that semantic theorem search is feasible at web scale. The project page is at \href{https://theoremsearch.com}{theoremsearch.com}. The dataset is available at \href{https://huggingface.co/datasets/uw-math-ai/theorem-search-dataset}{huggingface.co/datasets/uw-math-ai/theorem-search-dataset}. A datasheet for our dataset is provided in Appendix~\ref{sec:datasheet}.

\section{Related Work}

\paragraph{Mathematical Information Retrieval.}
Early MathIR work focused on formula-level retrieval, with the NTCIR Math Tasks \cite{aizawa2014ntcir,zanibbi2016ntcir} establishing benchmarks for formula search over arXiv and Wikipedia. The ARQMath shared tasks \cite{Mansouri2022OverviewOA} extended this to mathematical question answering over Math Stack Exchange. A recent survey \cite{dadure2024mathir} notes that while formula retrieval has progressed, semantic understanding of mathematical statements remains an open challenge -- the gap our work addresses.

\paragraph{Dense Retrieval and LLM-Augmented Search.}
Dense Passage Retrieval \cite{karpukhin2020dense} showed that dual-encoder architectures can outperform sparse methods like BM25. Sentence-BERT \cite{reimers2019sentence} enabled efficient semantic similarity via Siamese networks, while E5 \cite{wang2022text}, Qwen3-Embedding \cite{qwen3embedding}, and Gemma Embedding \cite{embedding_gemma_2025} have pushed embedding quality further. ColBERT \cite{khattab2020colbert} introduced late interaction for fine-grained token matching. Retrieval-augmented generation (RAG) \cite{lewis2020rag} combines retrieval with language model generation, enabling systems to ground responses in retrieved documents. These advances underpin our approach, though mathematical text poses unique challenges due to symbolic notation.

\paragraph{Search for Formal Mathematics.}
LeanSearch \cite{gao2025semanticsearchenginemathlib4} provides semantic search over Mathlib4's 230,000+ theorems by generating natural-language descriptions and using dense retrieval. ReProver \cite{yang2023leandojo} uses retrieval-augmented generation to select premises during proof search, and Numina-Lean-Agent \cite{liu2026numina} integrates multiple search tools, including LeanDex, to retrieve lemmas across libraries. LeanFinder \cite{lu2025leanfinder} focuses on user intent, while LeanExplore \cite{asher2025leanexplore} combines embeddings with BM25+ and PageRank. \citet{jiang2023multilingualmathematicalautoformalization} demonstrate that language models can translate between formal and natural language mathematics. However, formal libraries cover only a fraction of mathematical knowledge; our work extends informalization-based retrieval to millions of \LaTeX{} theorem statements.

\paragraph{Scientific Literature Search.}
General academic search engines (Google Scholar, Semantic Scholar) and arXiv provide paper-level retrieval but cannot target individual theorems. The zbMATH database \cite{zbmath} offers curated paper-level indexing. Large language models with web access, such as GPT-4 \cite{GPT4} and Gemini \cite{Gemini3}, can answer mathematical questions and sometimes locate relevant papers, but as we show in our experiments, they often provide incorrect theorem references or fail to locate specific statements. Our work treats theorem statements as first-class retrieval objects, enabling users to find specific results rather than papers that might contain them.

\section{Data Collection}

\begin{figure*}
    \centering
    \includegraphics[width=1.0\linewidth]{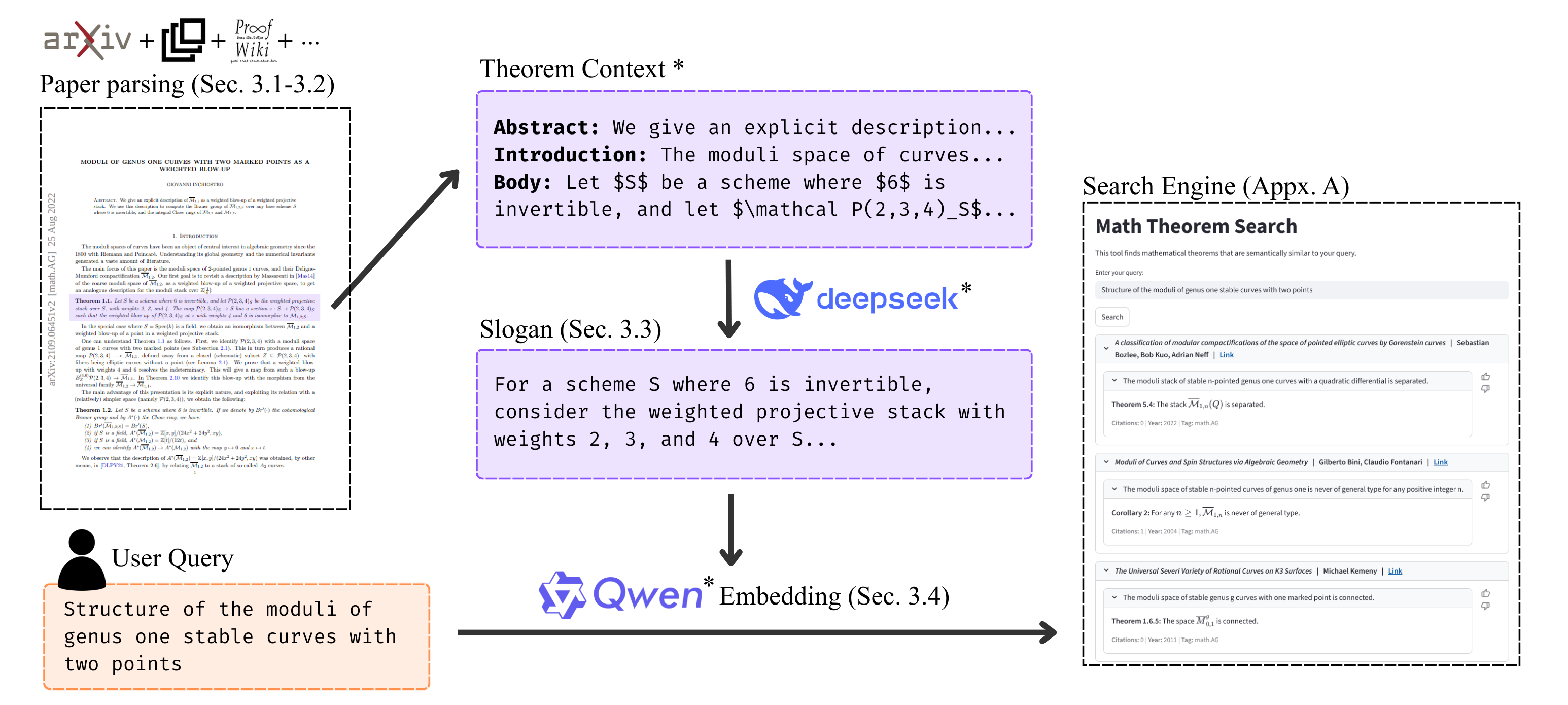}
    \caption{Overview of the creation of our database and search engine. Items with a ``*'' are variable and we tested different methods during our experiments (Sec. 4).}
    \label{fig:mts-overview}
\end{figure*}

Figure~\cref{fig:mts-overview} gives an overview of the pipeline used to parse theorem statements from source documents, generate natural-language representations, and embed them for retrieval.

\begin{figure}
    \centering
    \includegraphics[width=\linewidth]{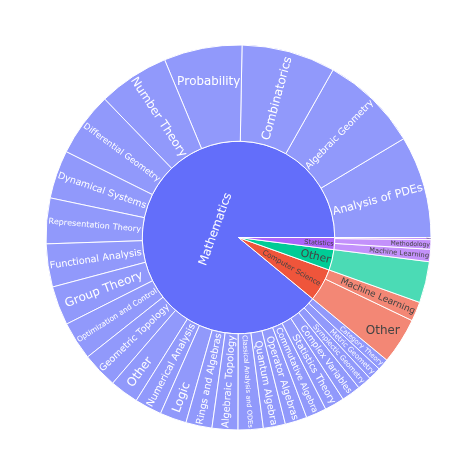}
    \caption{Sunburst plot of the distribution of arXiv tags across theorems in our dataset.}
    \label{fig:theorems_sunburst}
\end{figure}

\begin{table}[h]
\caption{Distribution of theorem types in our dataset.}
\label{tab:theorem-types}
\vskip 0.15in
\centering
\footnotesize
\setlength{\tabcolsep}{6pt}
\renewcommand{\arraystretch}{1.2}
\begin{tabular}{lc}
\toprule
\textbf{Theorem Type} & \textbf{Count} \\
\midrule
Lemma & 3,280,463 \\
Theorem & 2,864,668 \\
Proposition & 2,067,371 \\
Corollary & 1,034,259 \\
\bottomrule
\end{tabular}
\vskip -0.1in
\end{table}

\subsection{Theorem Parsing from arXiv}
We source 99.5\% (9.2 million) of our theorems from papers in the arXiv database tagged with \verb|math|, \verb|stat|, \verb|cs|, \verb|physics|, \verb|eess|, \verb|econ|, \verb|q-fin|, or \verb|q-bio|. A full breakdown of the statement types and arXiv categories across the dataset are provided in Table~\cref{tab:theorem-types}, Figure~\cref{fig:theorems_sunburst}, and Table~\ref{tab:arxiv-tags-full} in Appendix~\cref{val-set-arxiv-categories}.

Our objective in parsing theorems is to extract theorem \textit{names} and \textit{bodies}. A theorem \textit{name} has three components:
\begin{enumerate}
    \item The theorem type (e.g., Theorem, Proposition, Lemma, or Corollary),
    \item An optional reference number, and 
    \item An optional note. For example, a theorem name might look like ``Theorem 3.9 (Shokurov reduction)''.
\end{enumerate}
We define the \textit{body} as the \LaTeX{} content of the theorem, in which basic author-defined macros are expanded. For instance, if an author defines \verb|\R| as a macro for \verb|\mathbb{R}| in the \LaTeX{} preamble, we replace all occurrences of \verb|\R| accordingly.

The variety of ways authors define theorem environments makes parsing theorems difficult. Thus, we parse theorems for each paper using three strategies:
\begin{enumerate}
    \item \textit{Node Search with plasTeX:} We use the Python library plasTeX \cite{Arnold2009plasTeX} to convert \LaTeX{} sources into a structured node tree. Every command and environment corresponds to a node. We traverse this tree to find environments corresponding to theorem-like structures. Then, we extract the theorem type, reference number, note, and body from the node metadata. This method successfully parses 422 thousand papers, yielding approximately 6.9 million theorems. However, plasTeX occasionally truncates theorem bodies when a paper’s source relies on \LaTeX{} packages unknown to the parser. To remove such malformed extractions, we apply simple heuristics, such as filtering out theorems shorter than 8 characters or ending in `` and'' or `` let''.
    \item \textit{TeX Logging:} As a fallback to plasTeX, we generate and inject a custom \LaTeX{} package that logs theorem data. When a paper’s source is compiled, the package records the theorem type, reference number, note, and body for all theorem environments. We parse 137 thousand papers this way, yielding approximately 1.8 million theorems.
    \item \textit{Regex-based Parsing:} As a fallback to TeX logging, we use regular expressions to identify theorem delimiter tokens such as \verb|\begin{...}| and \verb|\end{...}| or \verb|\proclaim| and \verb|\endproclaim|. We then parse the content between delimiters to extract the theorem’s note and body. Reference numbers are not captured unless they are explicitly stated. We parse 30,000 papers this way, yielding approximately 542,000 theorems. 
\end{enumerate} 

\subsection{Theorem Parsing from Other Sources}
The remaining 0.5\% (38,974) of our theorems come from a variety of other sources: ProofWiki \cite{ProofWiki} (23,871), the Stacks Project \cite{StacksProject} (12,693), the Open Logic Project \cite{OpenLogicProject} (745), the CRing Project \cite{CRingProject} (546), Stacks and Moduli \cite{StacksandModuli} (506), the HoTT Book \cite{HoTTBook2013} (382), and An Infinitely Large Napkin \cite{ChenNapkin} (231). These sources were chosen for their verifiability through open-source contributions and structured, accessible \LaTeX{} files. We favored graduate-level texts, as their theorems would be useful to a mathematician and are often cited by researchers.

The Stacks Project, Open Logic Project, CRing Project, HoTT Book, and Infinitely Large Napkin are all hosted as open-source repositories on GitHub. Theorems from these sources were exclusively delimited in the source by \verb|\begin{...}| and \verb|\end{...}| tags. This uniform structure allowed us to build a single regex-based parser, similar to our regex-based parser for arXiv, that could extract theorems from all five sources. Our parser first normalizes shorthand environment names, then extracts the theorem name, body, label, and type from each environment. When a theorem body begins with a bracketed note, we append it to the theorem name in parentheses for easier identification. Labels defined with \verb|\label{...}| are captured to preserve cross-referencing information from the original text. Since these repositories use GitHub's file structure, we generate source URLs pointing to the original \texttt{.tex} files, enabling users to verify theorems against their source material. In addition, a custom theorem counter was used to reliably generate reference numbers.

ProofWiki required a different approach, as it is a wiki-based encyclopedia built on MediaWiki rather than \LaTeX{}. ProofWiki stores content in wikitext format with mathematical expressions enclosed in \verb|<math>...</math>| tags. We developed a separate parser that interfaces with the MediaWiki API to programmatically retrieve pages from relevant categories. For each page, we extract the theorem statement by identifying section headers and capturing content up to the proof section. The wikitext is then cleaned by removing MediaWiki-specific markup such as \verb| <onlyinclude>| tags, template calls, and wiki links, while converting \verb|<math>| blocks to standard delimiters. Since ProofWiki assigns a unique URL to each theorem, we preserve these links to allow direct verification.

\subsection{Theorem Representation}
Theorems in our corpus are exclusively represented in \LaTeX{}, and often lack concise natural-language summaries. To obtain searchable textual representations and improve retrieval performance, we generate a short natural language description, or \emph{slogan}, for each theorem using a large language model (LLM). This converts the retrieval task from a symmetric search over formal notation to an asymmetric task where informal queries retrieve formalized content \cite{wang2022text}. Given the parsed theorem body, we prompt the DeepSeek V3 model \cite{DeepSeekV3} to produce a concise, declarative English description of the theorem's main result. Prompts instruct the LLM to avoid symbolic notation, proof details, and references to surrounding document structure in the generated slogans. The resulting slogans are stored as the primary textual representation for theorem retrieval.

To study the effect of additional mathematical context on slogan quality and retrieval performance, we evaluate three slogan-generation strategies:
\begin{enumerate}
    \item \textbf{Body Only}: The prompt includes only the parsed theorem body.
    \item \textbf{Body + Abstract}: The prompt includes the theorem body together with the paper abstract.
    \item \textbf{Body + Introduction}: The prompt includes the theorem body together with the paper introduction.
\end{enumerate}

Each strategy uses a similar prompt template as shown in \cref{slogan-prompts}. Their major differences are in the contextual fields provided to the model. Model temperature is fixed at 0.2, and the maximum output tokens is fixed at 1024. Slogans are generated independently for each variant and treated as separate representations of the same underlying theorem during evaluation. The total cost of building the corpus was approximately \$6{,}000 USD: \$4{,}000 for LLM API calls to generate slogans across all 9.2 million theorems, and \$2{,}000 for compute and S3 storage.

\subsection{Result Retrieval}

We embed theorem slogans and user queries using the Qwen3-Embedding-8B model \cite{qwen3embedding}, which maps natural language inputs to fixed-dimensional vectors. All theorem slogans are embedded offline following their generation by the LLM and stored in a PostgreSQL database \cite{StonebrakerRowe1986Postgres} with the pgvector extension \cite{pgvector}. The database employs a Hierarchical Navigable Small World (HNSW) index \cite{malkov2018efficientrobustapproximatenearest} combined with binary quantization, enabling fast approximate nearest-neighbor search. User queries are embedded at inference time using the same method. Within this shared embedding space, we retrieve the top-$k$ theorems ranked by their Hamming distance, then re-ranked by cosine similarity.

\subsection{Validation Set}
Enlisting the help of three research mathematicians, we curated a set of 111 distinct math queries across 14 arXiv tags, mainly Algebraic Geometry, Analysis, and PDEs. See Table \ref{tab:arxiv-tags} in the Appendix for a detailed breakdown. These queries search for distinct theorems, lemmas, corollaries, and propositions written by a small number of authors whose work our subject matter experts are well-acquainted with. Crucially, queries were written \emph{blind}: each mathematician composed natural-language descriptions of theorems they knew from memory, without access to the corpus or its slogans, to prevent any leakage of retrieval representations into the evaluation queries. We then asked them to select a larger dataset of papers in which they were confident that the results of the queries they wrote appeared at most once; this larger dataset consisted of 7,356 papers. We used this larger dataset to guide our decisions on which embedders to use, which LLM to use for the slogans, how to prompt the embedder, and how much context to give to the LLM when generating the slogans (body of the result only, body+abstract, body+first section). Every query--theorem pair was subsequently verified in a two-stage quality check: an LLM confirmed that the target theorem exists in the corpus and semantically matches the query, and a second mathematician independently reviewed each pair to ensure correctness.

We remark that while small, this size is typical of mathematical information retrieval datasets, such as the query set of LeanSearch for Mathlib4 \cite{gao2025semanticsearchenginemathlib4} and ARQMath-3 \cite{Mansouri2022OverviewOA}, which contain 50 and 78 queries, respectively.

\section{Experiments}

In this section, we report the evaluation performance of existing retrieval methods against our search engine. We outline the experiment setup, compare the performance against existing retrieval tools, and perform an ablation study on both the context window and the LLM used for slogan generation.

\subsection{Metrics}

To measure the performance of our search engine against existing literature review tools, we employ two commonly used information retrieval metrics, following standard practice in mathematical retrieval benchmarks \cite{Mansouri2022OverviewOA, dadure2024mathir}. Hit@$k$, also known as the hit rate, measures whether the correct result appears anywhere in the top $k$ results:
\[
\text{Hit}@k = \frac{1}{|\mathcal{Q}|}\sum_{i=1}^{|\mathcal{Q}|} \max_{1 \le j \le k}\, \mathbb{I}(r_{i,j})
\]
where $\mathcal{Q}$ is the set of queries, $r_{i,j}$ is the $j$-th result of the $i$-th query, and $\mathbb{I}(r_{i,j})$ equals 1 if and only if $r_{i,j}$ is an exact match. We also use Mean Reciprocal Rank (MRR@$k$) \cite{Craswell2009}, which takes the reciprocal of the rank at which an exact match is discovered:
\[
\text{MRR}@k = \frac{1}{|\mathcal{Q}|}\sum_{i=1}^{|\mathcal{Q}|}  \frac{1}{\text{rank}_i}
\]
where $\text{rank}_i$ is the position of the first relevant result.

\subsection{Setup}


We chose three embedders to use in our experiments: Google's Gemma 0.3B \cite{embedding_gemma_2025}, and both Qwen3 0.6B and 8B \cite{qwen3embedding}. Our baseline methods consist of the following: Filtered Google Search ($\texttt{site:arxiv.org}$ + Query), arXiv advanced search, ChatGPT 5.2 w/ Search, and Gemini 3 Pro \cite{Gemini3}; the prompts used for the LLM baselines are provided in Table~\ref{retrieval-prompts} in the Appendix. 

Both Google and arXiv are unable to return specific math statements, so their performance is graded on their ability to locate the correct paper. Conversely, LLM-based systems like ChatGPT and Gemini occasionally return correct results with incorrect reference numbers; we grade these as misses on theorem-level retrieval and treat them as paper matches for paper-level retrieval.

We emphasize that these baselines are not designed for theorem-level retrieval, but we evaluate them to reflect the tools currently used by LLMs and research mathematicians in the absence of specialized theorem-level search systems.

\subsection{Main Results}

\begin{table*}
\caption{Results on the validation set, comparing embedder performance against existing literature search tools. Values are reported as \textcolor{blue}{theorem-level} / \textcolor{red}{paper-level}. Google Search and arXiv search cannot return specific theorem statements, so only paper-level results are reported. Our corpus was restricted to arXiv papers for evaluation.}
\label{embedder-table}
\vskip 0.15in
\begin{center}
\begin{small}
\begin{sc}
\begin{tabular}{lcccc}
\toprule
Model & Hit@1 & Hit@10 & Hit@20 & MRR@20 \\
\midrule
\multicolumn{5}{c}{Baseline}\\
\midrule
Google Search\textsuperscript{\textdagger} & \textcolor{red}{0.162} & \textcolor{red}{0.378} & \textcolor{red}{0.378} & \textcolor{red}{0.237} \\
arXiv Search\textsuperscript{\textdagger} & \textcolor{red}{0.009} & \textcolor{red}{0.018} & \textcolor{red}{0.027} & \textcolor{red}{0.011} \\
Chat-GPT 5.2 & \textcolor{blue}{0.117} & \textcolor{blue}{0.180} & \textcolor{blue}{0.198} & \textcolor{blue}{0.139} \\
Gemini 3 Pro & \textcolor{blue}{0.171} & \textcolor{blue}{0.252} & \textcolor{blue}{0.270} & \textcolor{blue}{0.196} \\

\midrule
\multicolumn{5}{c}{Our Methods}\\
\midrule
Gemma 0.3B (\LaTeX) & \textcolor{blue}{0.027} / \textcolor{red}{0.054} & \textcolor{blue}{0.090} / \textcolor{red}{0.117} & \textcolor{blue}{0.090} / \textcolor{red}{0.135} & \textcolor{blue}{0.041} / \textcolor{red}{0.070} \\
Gemma 0.3B & \textcolor{blue}{0.081} / \textcolor{red}{0.108} & \textcolor{blue}{0.189} / \textcolor{red}{0.225} & \textcolor{blue}{0.252} / \textcolor{red}{0.306} & \textcolor{blue}{0.118} / \textcolor{red}{0.154} \\
Qwen3 0.6B & \textcolor{blue}{0.081} / \textcolor{red}{0.153} & \textcolor{blue}{0.234} / \textcolor{red}{0.342} & \textcolor{blue}{0.270} / \textcolor{red}{0.351} & \textcolor{blue}{0.132} / \textcolor{red}{0.215} \\
Qwen3 8B & \textcolor{blue}{0.171} / \textcolor{red}{0.243} & \textcolor{blue}{0.387} / \textcolor{red}{0.505} & \textbf{\textcolor{blue}{0.450}} / \textcolor{red}{0.568} & \textcolor{blue}{0.243} / \textcolor{red}{0.328} \\
Qwen3 8B w/ Reranker & \textbf{\textcolor{blue}{0.189}} / \textbf{\textcolor{red}{0.324}} & \textbf{\textcolor{blue}{0.432}} / \textbf{\textcolor{red}{0.613}} & \textbf{\textcolor{blue}{0.450}} / \textbf{\textcolor{red}{0.631}} & \textbf{\textcolor{blue}{0.270}} / \textbf{\textcolor{red}{0.416}} \\
\bottomrule
\multicolumn{5}{l}{\textsuperscript{\textdagger}\scriptsize Paper-level only: these tools cannot return individual theorems.}
\end{tabular}
\end{sc}
\end{small}
\end{center}
\vskip -0.1in
\end{table*}

Our results are presented in Table~\cref{embedder-table}. We find that the embedder Qwen3 8B outperforms existing literature review methods across all metrics. In particular, Qwen3 8B achieves substantially higher Hit@10 and Hit@20, indicating that the correct theorem is frequently retrieved within the top candidate set even when it is not ranked first. This behavior is desirable in large-scale theorem retrieval, where downstream reranking or human inspection can refine results once relevant candidates have been surfaced. We further apply a cross-encoder reranker (Qwen3-Reranker-0.6B \cite{qwen3embedding}) to rescore the top-100 candidates retrieved by Qwen3 8B; this improves theorem-level Hit@1 from 17.1\% to 18.9\% and MRR@20 from 24.3\% to 27.0\%, confirming that late interaction over full query--slogan pairs captures fine-grained semantic distinctions that the bi-encoder alone misses.

For search methods at the paper-level, Qwen3 8B continues to outperform the baseline models, achieving higher Hit@1 and MRR. The arXiv search is limited in the fields it can search over and does not search within the paper body, so it was only able to locate two papers across the whole query set in the top-20 results.

One notable effect we observed concerns how Chat-GPT and Gemini rank retrieved results. Although both models used web-search-based retrieval, they would often organize their outputs primarily at the paper level, returning multiple theorems from the same paper in consecutive positions before moving to another paper. This ranking pattern reduces result diversity, which lowers Hit@$k$ despite sometimes achieving high early precision. On average, Gemini returns only 10.98 distinct papers per query, compared to 16.89 for Qwen3 8B, indicating that our retrieval pipeline surfaces a broader cross-section of the literature.

Our approach is particularly advantageous for retrieving auxiliary lemmas and technical results that appear deep within a paper, far from the title and abstract. Google Search and LLM-based tools rely heavily on paper-level metadata and thus struggle to surface results whose content is not reflected in the abstract, whereas our system indexes every theorem independently via its slogan, making it equally as capable of retrieving a minor lemma in a later section as a paper's headline theorem.

\subsection{Retrieval-Augmented LLM Reasoning}
\label{sec:claude-main}

Beyond quantitative retrieval metrics, we investigate whether our theorem corpus can improve LLM reasoning on research-level mathematical questions. We posed a question about the boundary of a KSBA compactification of elliptic surface pairs to Claude (Opus 4.5). Without retrieval augmentation, Claude produced a confidently stated but \emph{incorrect} answer, concluding that all boundary pairs must retain a fibration structure. The argument, while superficially plausible, relied on incorrect claims about the behavior of slc surface pairs under KSBA compactification.

When the same question was posed with access to our theorem database as a RAG tool, Claude retrieved relevant results from \cite{ascher2017logcanonicalmodelselliptic, ascher2018moduliweightedstableelliptic, inchiostro2018moduliweierstrassfibrationsmarked, ascher2016modulifiberedsurfacepairs} and arrived at the \emph{correct} answer: such boundary pairs do exist, and they are pseudoelliptic surfaces obtained by contracting the section of an elliptic fibration. The RAG-augmented response cited specific theorem numbers and provided a logically valid chain of reasoning grounded in retrieved results. This illustrates a key failure mode of LLM reasoning---without access to the relevant literature, the model confabulates plausible-sounding but incorrect mathematical arguments---and demonstrates that retrieval augmentation over our corpus can directly address it. The full query, both responses, and detailed citations are reported in Appendix~\ref{sec:claude_work}.

\subsection{Ablation Studies}

In this subsection, we perform ablation studies on the context windows used for slogan generation, the LLMs used to generate slogans, and the document preparation process. For each ablation, we limit the corpus to arXiv papers labeled with the algebraic geometry tag (math.AG) and authored by individuals whose work contains an exact match to one of our queries. This filtering reduces the corpus to 7,356 statements written by 8 primary authors.

\subsubsection{Ablation of Context Windows}

Table~\Cref{context-window-table} presents the effects of context on slogan generation. Increasing or decreasing the amount of context provided to LLMs when generating math slogans considerably affects our search engine's performance, while holding the embedder constant. We find that supplying the LLM with additional paper context improves the retrieval performance of its generated slogans. Performance falters when the LLM is only given the abstract, but improves significantly with the inclusion of the first section of the paper, defined by \texttt{\textbackslash section\{\}}. We attribute this improvement to the LLM making better sense of the statement with more context, especially since theorems alone tend to be at most 3--4 sentences long and may reference earlier sections of the paper. Furthermore, the introduction of most math papers typically outlines the results and the existing research upon which they build. As a result, context-rich slogans better capture the semantic intent of the theorem, yielding more reliable and robust retrieval across queries. Results for an expanded ablation covering additional embedders are reported in Table~\ref{context-window-table-full} in Appendix~\ref{more-experiments}.

\begin{table}[ht]
\caption{Comparing context window size in retrieval performance. Embedded with prompt using Qwen3 8B \cite{qwen3embedding}. Slogans generated with DeepSeek V3 \cite{DeepSeekV3}.}
\label{context-window-table}
\vskip 0.15in
\begin{center}
\begin{small}
\begin{sc}
\resizebox{\columnwidth}{!}{
\begin{tabular}{lcccc}
\toprule
Model & Hit@1 & Hit@10 & Hit@20 & MRR@20 \\
\midrule
\multicolumn{5}{c}{Context Windows}\\
\midrule
Body Only & 0.342 & 0.658 & 0.737 & 0.451 \\
w/ Abstract & 0.316 & 0.645 & 0.737 & 0.429 \\
w/ First Section & \textbf{0.368} & \textbf{0.737} & \textbf{0.763} & \textbf{0.496} \\
\bottomrule
\end{tabular}
}
\end{sc}
\end{small}
\end{center}
\vskip -0.1in
\end{table}

\subsubsection{Ablation of Slogan Generator}

The LLM used to generate the slogans can also significantly affect our engine's search performance, as some LLMs are better trained on mathematical texts than others and thus generate better informal versions of math statements. We find that leading proprietary models, such as Claude's Opus 4.5 and Gemini 3 Pro \cite{Gemini3}, outperform open-source models such as DeepSeek-V3 \cite{DeepSeekV3}, as summarized in Table~\cref{slogan-llm-table}. A broader comparison across additional slogan generators and embedding models is provided in Table~\ref{slogan-llm-table-full} in Appendix~\ref{more-experiments}.

\begin{table}[h]
\caption{Comparing LLM slogans in retrieval performance. Embedded with prompt using Qwen3 8B \cite{qwen3embedding}. Body + Abstract}
\label{slogan-llm-table}
\vskip 0.15in
\begin{center}
\begin{small}
\begin{sc}
\resizebox{\columnwidth}{!}{
\begin{tabular}{lcccc}
\toprule
Model & Hit@1 & Hit@10 & Hit@20 & MRR@20 \\
\midrule
\multicolumn{5}{c}{LLMs}\\
\midrule
Deepseek V3.1 & 0.316 & 0.645 & 0.737 & 0.429 \\
Deepseek R1 & 0.276 & 0.671 & 0.697 & 0.388 \\
Gemini 3 Pro & 0.368 & 0.750 & 0.816 & 0.507 \\
Claude Opus 4.5 & \textbf{0.395} & \textbf{0.776} & \textbf{0.842} & \textbf{0.536} \\
\bottomrule
\end{tabular}
}
\end{sc}
\end{small}
\end{center}
\vskip -0.1in
\end{table}

\subsubsection{Ablation of Task Instruction}

\begin{table}
\caption{Comparing prompting performance. Slogans generated by DeepSeek V3.1 \cite{DeepSeekV3} on `Body+Abstract' context window.}
\label{doc-table-prep}
\vskip 0.15in
\begin{center}
\begin{small}
\begin{sc}
\resizebox{\columnwidth}{!}{
\begin{tabular}{lcccc}
\toprule
Model & Hit@1 & Hit@10 & Hit@20 & MRR@20 \\
\midrule
\multicolumn{5}{c}{Unprompted}\\
\midrule
Gemma 0.3B & 0.303 & 0.539 & 0.566 & 0.376 \\
Qwen3 0.6B & 0.224 & 0.487 & 0.526 & 0.297 \\
Qwen3 8B & 0.250 & 0.553 & 0.618 & 0.332 \\
\midrule
\multicolumn{5}{c}{Prompted}\\
\midrule
Gemma 0.3B & 0.197 & 0.434 & 0.487 & 0.265 \\
Qwen3 0.6B & 0.237 & 0.566 & 0.658 & 0.346 \\
Qwen3 8B & \textbf{0.316} & \textbf{0.645} & \textbf{0.737} & \textbf{0.496} \\
\bottomrule
\end{tabular}
}
\end{sc}
\end{small}
\end{center}
\vskip -0.1in
\end{table}

We examine the impact of the task instructions used during embeddings on retrieval performance in Table~\cref{doc-table-prep}. We find that by providing a math retrieval instruction, both Qwen embedders achieve higher performance than without any instruction at all. However, this effect is reversed with Gemma, with empty instruction performing better, though worse than Qwen 8B with instruction. We use a modification of the prompts made by \cite{gao2025semanticsearchenginemathlib4}, the details of which are listed in Table~\cref{doc-prep-table-describe}.

\subsection{Embedding Space Analysis}

Following \cite{gao2025semanticsearchenginemathlib4, jiang2023multilingualmathematicalautoformalization}, we convert \LaTeX{} theorem statements into natural-language slogans before embedding, since embedders struggle with symbol-heavy notation \cite{bleckmann2025evaluatingnlpembeddingmodels, peng2021mathbertpretrainedmodelmathematical} and mathematicians typically query in informal language. To verify that sloganization induces semantically meaningful structure, we use PCA and UMAP as diagnostic tools on a random sample of 10,000 theorems (1,000 from each of the 10 most common arXiv categories).

\begin{figure}[b]
\centering
\includegraphics[width=0.49\linewidth]{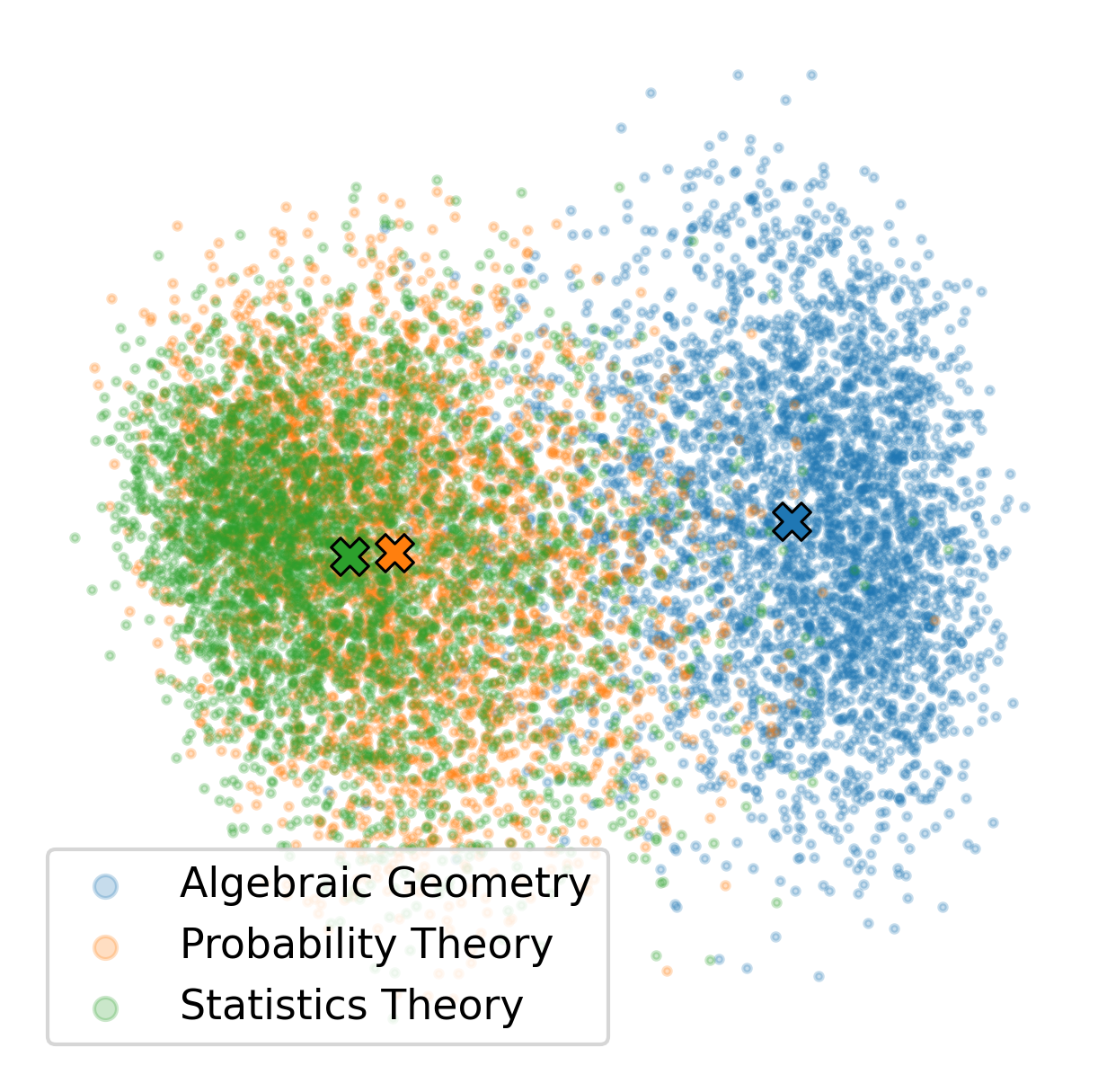}\hfill
\includegraphics[width=0.49\linewidth]{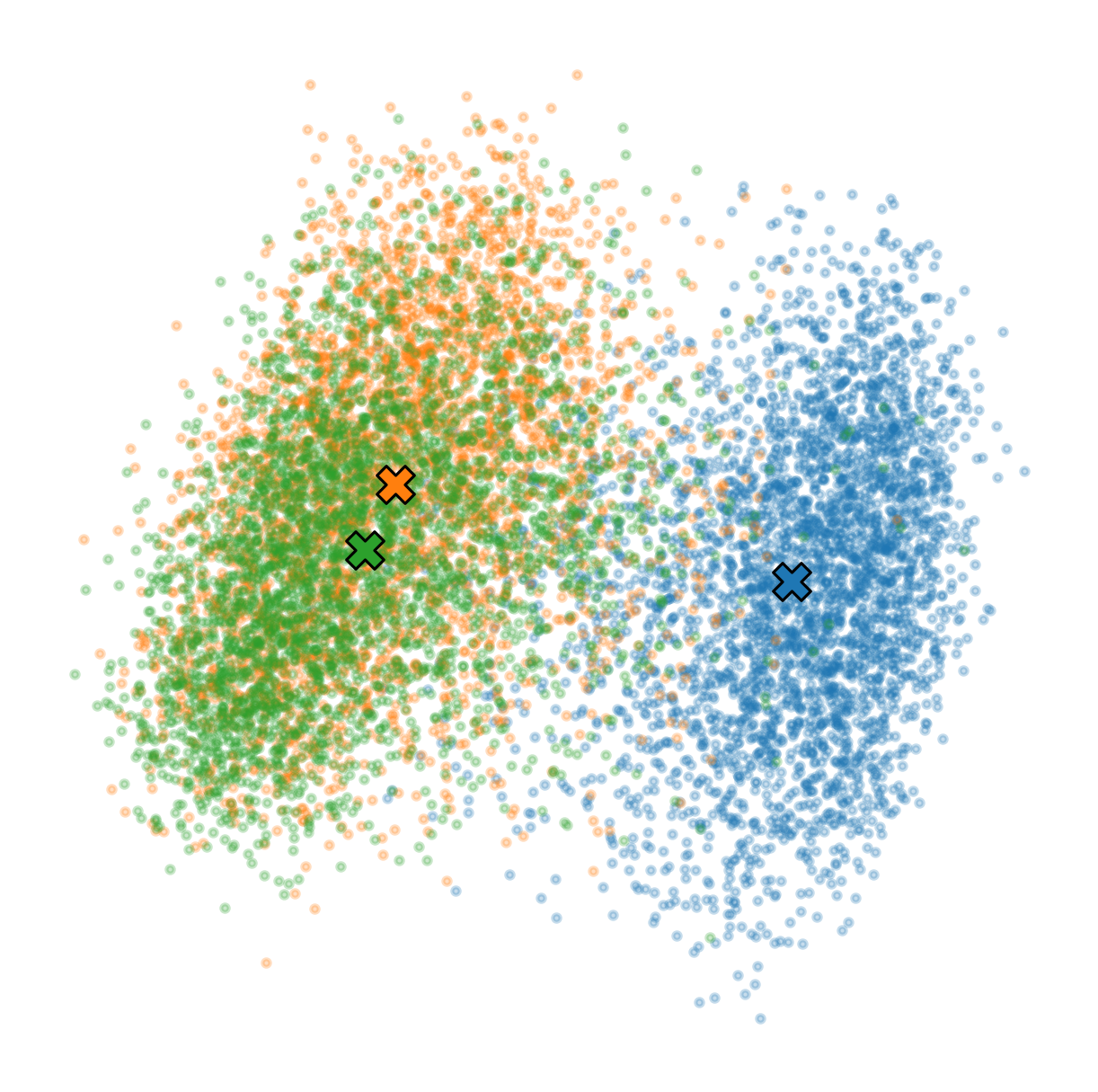}
\caption{PCA visualizations of theorems in our dataset using slogan embeddings.
Left: Gemma embedding. Right: Qwen3 8B embedding.}
\label{fig:pca-three-categories}
\end{figure}

\begin{figure*}
    \centering
    \includegraphics[width=0.36\linewidth]{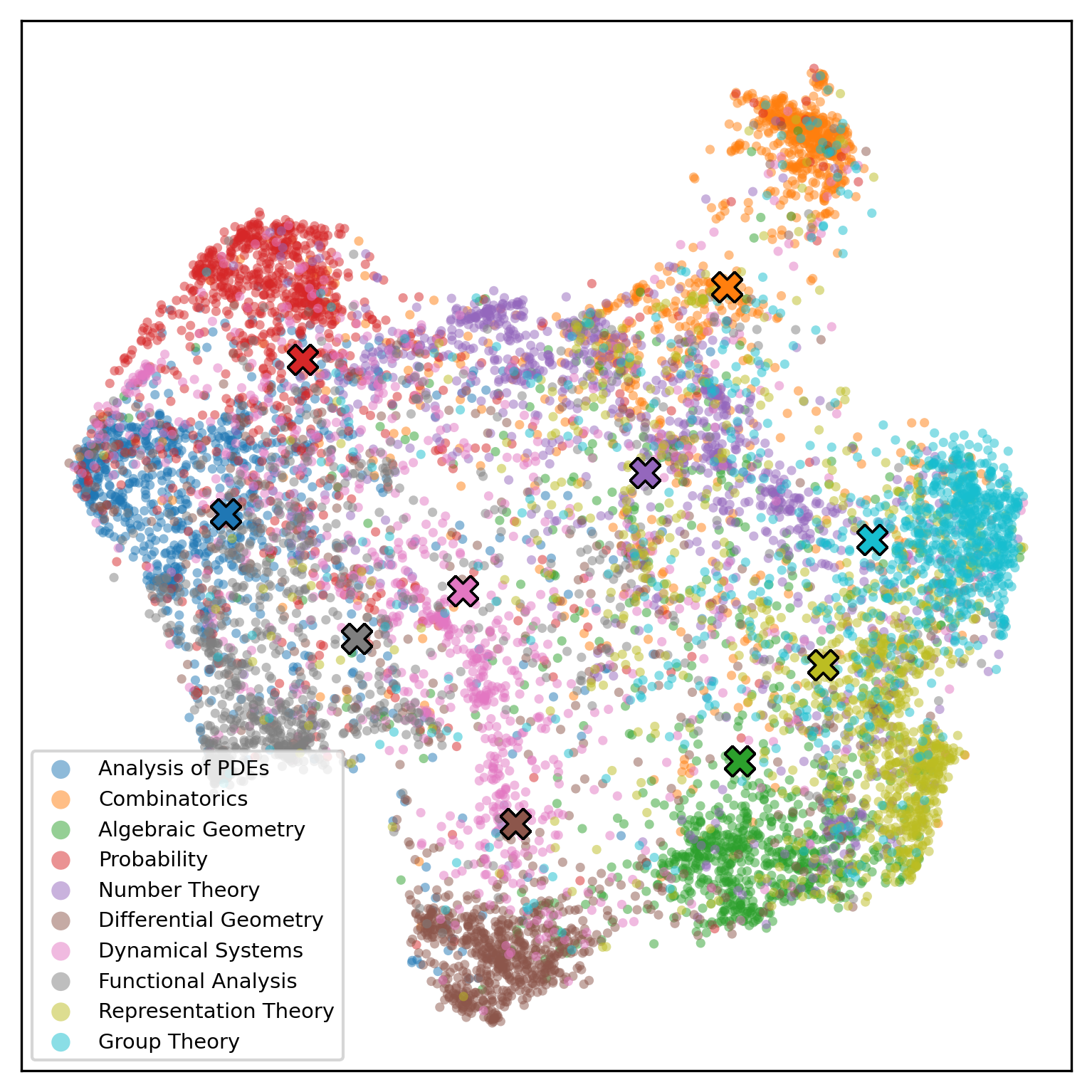}
    \includegraphics[width=0.36\linewidth]{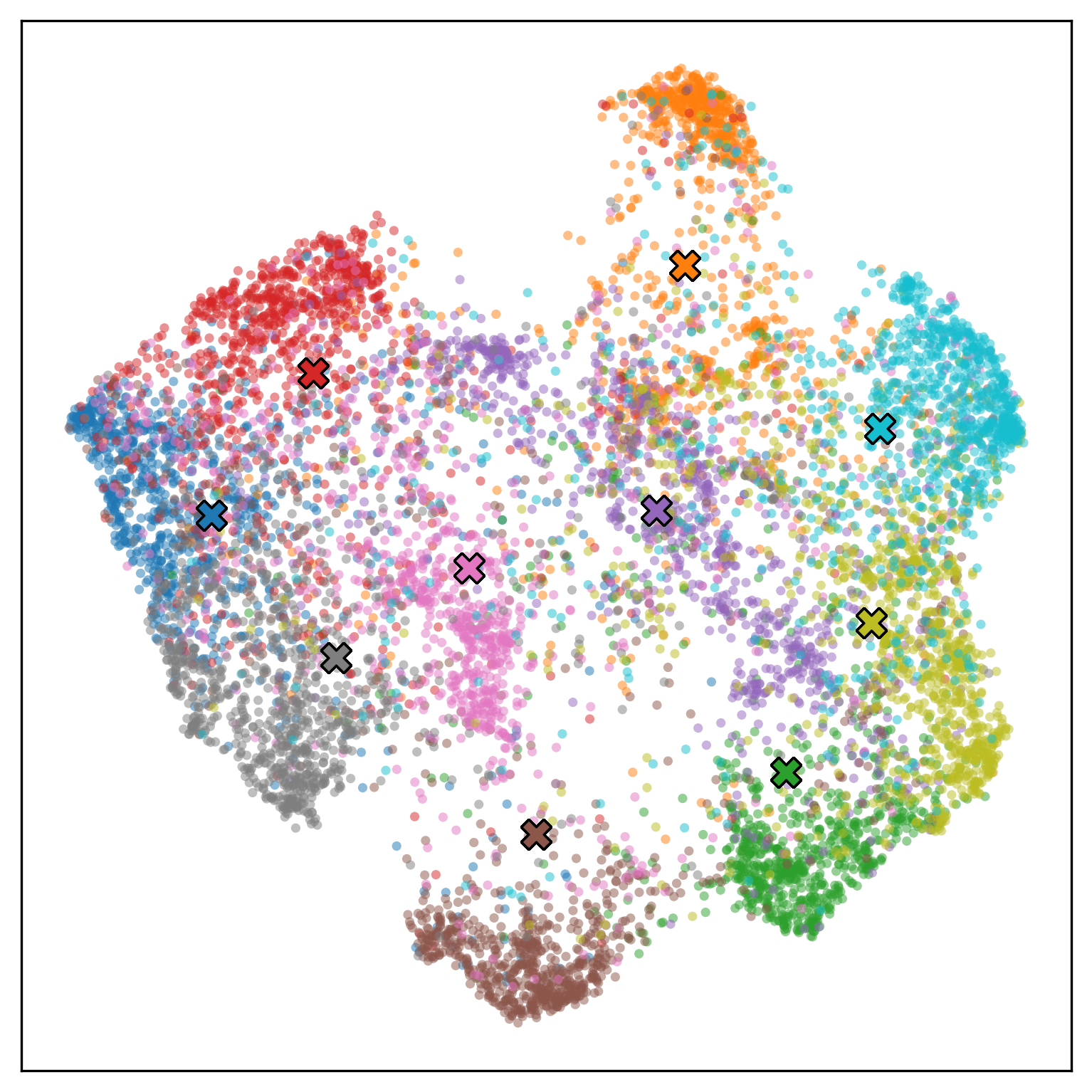}
    \caption{UMAP visualizations of 10,000 theorem slogan embeddings across the ten most common arXiv categories. Gemma 0.3B (left) and Qwen3 8B (right). Qwen3 8B produces tighter, better-separated clusters.}
    \label{fig:umap}
\end{figure*}

Figure~\cref{fig:pca-three-categories} shows PCA projections for three categories: Algebraic Geometry, Probability Theory, and Statistics Theory. Conceptually distant fields (Algebraic Geometry vs.\ Probability) are well separated, while closely related fields (Probability vs.\ Statistics) overlap substantially -- consistent with mathematical intuition. UMAP projections over all ten categories in Figure~\ref{fig:umap} reveal that Qwen3 8B produces tighter, better-separated clusters than Gemma 0.3B, consistent with the retrieval gap in Table~\ref{embedder-table}.

\section{Search Tool}

\label{sec:searchtool}

\begin{figure*}
    \centering
    \includegraphics[width=\linewidth]{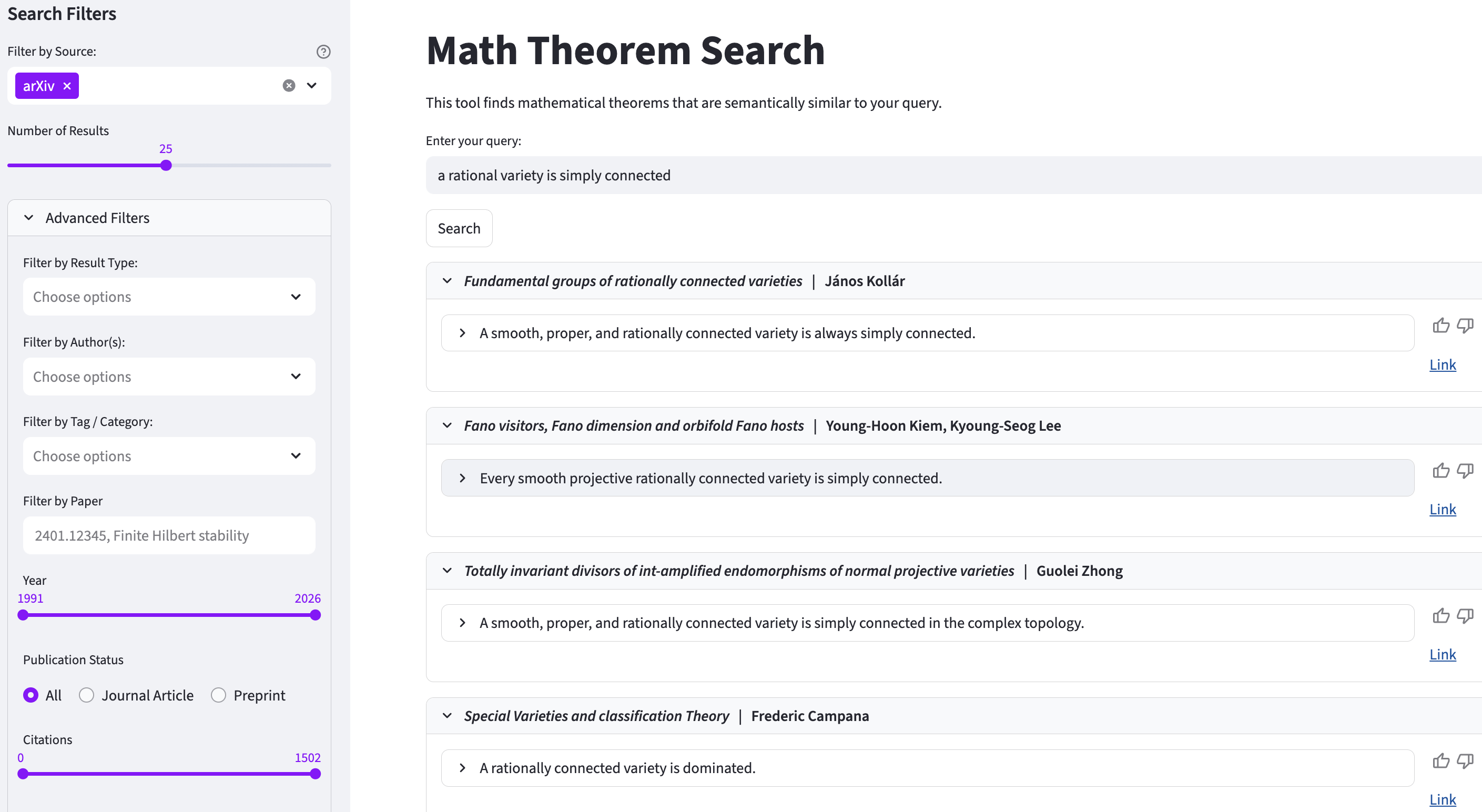}
    \caption{The interface of the Theorem Search Tool. The main window displays the top results: paper name, paper authors, and theorem statement, and link to the paper. In the left pane users can filter by paper metadata: type of result (theorem/proposition/lemma/corollary), author(s), arXiv tag, specific paper, year, and publication status. The thumbs up/down allow for user feedback, used to further improve the search results.}
    \label{fig:theorem-search-app}
\end{figure*}

The search tool is currently accessible via HuggingFace Spaces \href{https://huggingface.co/spaces/uw-math-ai/theorem-search}{(Link)}. Users can perform semantic searches using natural language (e.g., “a rational variety is simply connected”). The sidebar interface allows results to be filtered by relevant metadata (e.g., author sets, arXiv category tags, and journal publication status). To optimize retrieval time, the search uses a two-stage retrieval architecture. At inference time, the natural language query is encoded using the Qwen3-8B embedding model and binary quantized into 4096 bits. An initial pool of $\text{clamp}(\max(200, 12 \times k), 200, 800 )$ candidates is returned using the HNSW index and Hamming distance, where $k$ is a user-specified number of results to return. The candidate pool is then filtered by metadata and reranked using cosine similarity on the original 4096-dimensional embeddings. When weighting by citations is enabled, the candidates are instead re-ranked using the following composite scoring function: $$\text{score} = \text{cosine similarity} + \lambda \cdot \log (\max(\text{citations, 1}))$$ where $\lambda$ is the user-specified citation-weight parameter. The interface retrieves the top $k$ results, each of which displays the theorem slogan, the theorem body rendered in LaTeX, any relevant paper metadata, and a link to the paper. Finally, a lightweight feedback system allows users to rate each result's relevance to their query. The resulting end-to-end latency is approximately 3 seconds per query, compared to 60+ seconds for LLM-based search tools such as ChatGPT and Gemini, which must perform multiple web searches and parse documents at inference time. See Figure \cref{fig:theorem-search-app} for the interface.

Because our slogans are multi-sentence descriptions, the embedding model produces vectors that are optimized for longer inputs. Short keyword-style queries (e.g., ``Landau equation'') can therefore be dominated by padding or generic context, leading to poor retrieval. We find that simply repeating a short query (e.g., ``Landau equation Landau equation'') substantially improves result quality by strengthening the signal of the key terms in the embedding. The search tool applies this duplication automatically when the query is below a length threshold.

In addition to the web interface, we provide a REST API at \href{https://api.theoremsearch.com/search}{\texttt{api.theoremsearch.com/search}} that accepts JSON queries and returns ranked theorem results, enabling programmatic integration into research workflows. We also expose a Model Context Protocol (MCP) server at \href{https://api.theoremsearch.com/mcp}{\texttt{api.theoremsearch.com/mcp}}, allowing AI agents and LLM-based tools to use theorem search as a retrieval tool during reasoning and proof generation. The project page, including links to the search tool, API, MCP server, and dataset, is available at \href{https://theoremsearch.com}{theoremsearch.com}.

\section{Conclusion}

Mathematical knowledge is organized around discrete results -- theorems, lemmas, propositions, and corollaries -- yet existing search tools operate at the document level, forcing researchers and AI agents alike to manually locate specific statements within papers. In this work, we addressed this gap by constructing a corpus of over 9.2 million theorem statements extracted from arXiv and seven other sources, constituting the largest unified collection of human-authored, research-level theorems to date.

We showed that representing theorems via natural-language slogans generated by an LLM, rather than embedding raw \LaTeX{} directly, yields substantially better retrieval performance. Our systematic ablations revealed that context matters: slogans generated with access to the paper's introduction outperform those generated from the theorem body alone, and the choice of both the slogan-generating LLM and the embedding model significantly affects downstream retrieval quality. On a curated evaluation set of 111 queries written by professional mathematicians, our best configuration achieves 45.0\% Hit@20 at the theorem level and 56.8\% Hit@20 at the paper level, outperforming ChatGPT~5.2 with search, Gemini~3 Pro, and Google Search.

By treating theorems as first-class retrieval objects and enriching them with natural-language descriptions, we enable new forms of access to mathematical knowledge -- whether for retrieval-augmented generation by LLMs, for premise selection in formal proof search, or for literature review by expert mathematicians. We have deployed a public search interface described in Appendix~\ref{sec:searchtool}. Qualitative user feedback from research mathematicians, reported in Appendix~\ref{sec:user-feedback-appendix}, further illustrates the practical utility of our system.


\section*{Acknowledgments}
We thank Michele Pernice and Dori Bejleri for contributing several entries to the validation set.
CPU and GPU computing, LLM inference and data storage were in part done using AWS credits from the UW eScience School and UW IT.
CPU and GPU computing was in part done using the UW Research Computing Club funded from the UW Student Technology Fee Committee.
GPU computing was in part done on UWIT's GPU cluster Tillicum.
GPU computing and model inference were in part done using Nebius and TokenFactory.

\newpage

\bibliography{refs}
\bibliographystyle{icml2026}

\newpage

\appendix
\onecolumn

\section{Datasheet}
\label{sec:datasheet}

This section is based on \textit{Datasheets for Datasets} by \cite{gebru2021datasheetsdatasets}.

\subsection{Motivation}

\begin{enumerate}
    \item For what purpose was the dataset created?
    \begin{itemize}
        \item We created this dataset to create the largest unified corpus of mathematical theorems. We use this dataset ourselves to create a theorem search engine.
    \end{itemize}
    \item Who created the dataset?
    \begin{itemize}
        \item This paper's co-authors collected theorems. All theorems were written by the authors of the theorems' papers and are cited in our database.
    \end{itemize}
    \item Who funded the creation of the dataset?
    \begin{itemize}
        \item Self-funded.
    \end{itemize}
\end{enumerate}

\subsection{Composition}

\begin{enumerate}
    \item What do the instances that comprise the dataset represent?
    \begin{itemize}
        \item Each instance of the dataset represents a mathematical statement derived from a paper, textbook, or repository online.
    \end{itemize}
    \item How many instances are there in total?
    \begin{itemize}
        \item As of writing, there are 9.2 million mathematical statements in the dataset.
    \end{itemize}
    \item Does the dataset contain all possible instances or is it a sample (not necessarily random) of instances from a larger set?
    \begin{itemize}
        \item The dataset contains a non-random sample of instances from available math corpora.
    \end{itemize}
    \item What data does each instance consist of?
    \begin{itemize}
        \item Each instance contains the name of the theorem as displayed in the source document (e.g.\ Theorem 1.1), the raw TeX body, the author-written label inside the body, and the paper it was sourced from (e.g.\ arXiv's convention of 1234.56789v1).
        \item In addition, we stored metadata from each source document. This data includes the title, the list of authors, the link to the document, the abstract, the journal the document was published in (if it was published), the primary and secondary categories, the number of citations (if available), and a tag of which source it came from (e.g.\ arXiv, Stacks Project).
    \end{itemize}
    \item Is there a label or target associated with each instance?
    \begin{itemize}
        \item Each theorem uses its name derived from the source document, and lists the document it came from, creating a distinct method of identification.
    \end{itemize}
    \item Is any information missing from individual instances?
    \begin{itemize}
        \item No.
    \end{itemize}
    \item Are relationships between individual instances made explicit?
    \begin{itemize}
        \item Each instance is numbered according to the conventions of the source document. They also contain metadata linking them to a specific paper or document.
    \end{itemize}
    \item Are there recommended data splits? (e.g., training, development/validation, testing)
    \begin{itemize}
        \item Not applicable.
    \end{itemize}
    \item Are there any errors, sources of noise, or redundancies in the dataset?
    \begin{itemize}
        \item Some statement bodies, titles, and labels may be fragmented, as our parsing methods are not applicable to every TeX document in our sources.
    \end{itemize}
    \item Is the dataset self-contained, or does it link to or otherwise rely on external resources?
    \begin{itemize}
        \item The data is derived from sources online.
    \end{itemize}
    \item Does the dataset contain data that might be considered confidential?
    \begin{itemize}
        \item No, all data is publicly available online.
    \end{itemize}
    \item Does the dataset contain data that, if viewed directly, might be offensive, insulting, threatening, or might otherwise cause anxiety?
    \begin{itemize}
        \item No.
    \end{itemize}
\end{enumerate}

\subsection{Collection Process}

\begin{enumerate}
    \item How was the data associated with each instance acquired?
    \begin{itemize}
        \item A LaTeX parser was used to parse all theorem names and bodies from sources.
    \end{itemize}
    \item What mechanisms or procedures were used to collect the data?
    \begin{itemize}
        \item Our LaTeX theorem parser used the Python library plasTeX to parse LaTeX documents, with a custom TeX logger and a pure-regex parser as fallbacks.
    \end{itemize}
    \item If the dataset is a sample from a larger set, what was the sampling strategy?
    \begin{itemize}
        \item Our dataset is not a sample. To the best of our knowledge, our dataset includes most mathematical theorems in arXiv. For the other sources, such as ProofWiki and the Stacks Project, we have included all theorems.
    \end{itemize}
    \item Who was involved in the data collection process, and how were they compensated?
    \begin{itemize}
        \item This paper's co-authors were all involved in designing and implementing the pipeline to collect theorems. The mathematicians who created the validation set are volunteers.
    \end{itemize}
    \item Over what timeframe was the data collected?
    \begin{itemize}
        \item Data was collected over five months, from September 2025 to January 2026.
    \end{itemize}
    \item Were any ethical review processes conducted?
    \begin{itemize}
        \item Not applicable.
    \end{itemize}
\end{enumerate}

\subsection{Preprocessing/cleaning/labeling}

\begin{enumerate}
    \item Was any preprocessing/cleaning/labeling of the data done?
    \begin{itemize}
        \item We preprocessed theorem bodies by removing unnecessary whitespace and expanding simple author-defined macros. We filtered out truncated or corrupted theorem bodies by employing several heuristics.
    \end{itemize}
    \item Was the "raw" data saved in addition to the preprocessed/cleaned/labeled data?
    \begin{itemize}
        \item No.
    \end{itemize}
    \item Is the software that was used to preprocess/clean/label the data available?
    \begin{itemize}
        \item All code used to preprocess, clean, and label our data is available on our GitHub repo \href{https://github.com/uw-math-ai/TheoremSearch}{(Link)}.
    \end{itemize}
\end{enumerate}

\subsection{Uses}

\begin{enumerate}
    \item Has the dataset been used for any tasks already?
    \begin{itemize}
        \item The dataset has been used to create a semantic search engine, available on HuggingFace Spaces \href{https://huggingface.co/spaces/uw-math-ai/theorem-search}{(Link)}.
    \end{itemize}
    \item Is there a repository that links to any or all papers or systems that use the dataset?
    \begin{itemize}
        \item Not applicable.
    \end{itemize}
    \item What (other) tasks could the dataset be used for?
    \begin{itemize}
        \item Our dataset of theorems is useful for mathematicians and for AI agents for automated theorem proving.
    \end{itemize}
    \item Is there anything about the composition of the dataset or the way it was collected and preprocessed/cleaned/labeled that might impact future uses?
    \begin{itemize}
        \item If authors of our theorem sources were to significantly change the content of the papers/textbooks/projects, our theorems would be outdated.
    \end{itemize}
    \item Are there tasks for which the dataset should not be used?
    \begin{itemize}
        \item No.
    \end{itemize}
\end{enumerate}

\subsection{Distribution}

\begin{enumerate}
    \item Will the dataset be distributed to third parties outside of the entity on behalf of which the dataset was created?
    \item How will the dataset be distributed?
    \begin{itemize}
        \item The dataset is publicly available on Hugging Face at \href{https://huggingface.co/datasets/uw-math-ai/theorem-search-dataset}{huggingface.co/datasets/uw-math-ai/theorem-search-dataset}. We release only theorems from arXiv papers that use permissive licenses (CC BY and CC0), as well as theorems from the seven other sources (ProofWiki, Stacks Project, Open Logic Project, CRing Project, Stacks and Moduli, HoTT Book, and An Infinitely Large Napkin). Theorems from arXiv papers with non-permissive licenses are excluded from the public release.
    \end{itemize}
    \item When will the dataset be distributed?
    \begin{itemize}
        \item The dataset is already publicly available.
    \end{itemize}
    \item Will the dataset be distributed under a copyright or other intellectual property (IP) license, and/or under applicable terms of use (ToU)?
    \begin{itemize}
        \item The released dataset inherits the licenses of the source materials: CC BY or CC0 for arXiv papers, and the respective licenses of the other sources.
    \end{itemize}
    \item Have any third parties imposed IP-based or other restrictions on the data associated with the instances?
    \begin{itemize}
        \item No.
    \end{itemize}
    \item Do any export controls or other regulatory restrictions apply to the dataset or individual instances?
    \begin{itemize}
        \item No.
    \end{itemize}
\end{enumerate}

\subsection{Maintenance}

\begin{enumerate}
    \item Who will be supporting/hosting/maintaining the dataset?
    \begin{itemize}
        \item The co-authors of this paper plan on continuing to build, host, and maintain this dataset.
    \end{itemize}
    \item How can the owner/curator/manager of the dataset be contacted?
    \begin{itemize}
        \item The co-authors of this paper own this dataset and can be contacted through Vasily Ilin at \href{mailto:vilin@uw.edu}{vilin@uw.edu}.
    \end{itemize}
    \item Will the dataset be updated?
    \begin{itemize}
        \item We plan on continuing to improve and grow the dataset in the future. For example, we plan on adding theorems from nLab and other open-source textbooks and projects.
    \end{itemize}
    \item  If the dataset relates to people, are there applicable limits on the retention of the data associated with the instances (e.g., were the individuals in question told that their data would be retained for a fixed period of time and then deleted)?
    \begin{itemize}
        \item Not applicable.
    \end{itemize}
    \item Will older versions of the dataset continue to be supported/hosted/maintained?
    \begin{itemize}
        \item Older versions of the dataset will be available for download indefinitely on Hugging Face. However, all support and maintenance efforts will be focused on the latest version of the dataset.
    \end{itemize}
    \item If others want to extend/augment/build on/contribute to the dataset, is there a mechanism for them to do so?
    \begin{itemize}
        \item Others can suggest sources to add to the dataset by contacting Vasily Ilin at \href{mailto:vilin@uw.edu}{vilin@uw.edu}.
    \end{itemize}
\end{enumerate}

\section{User Feedback}
\label{sec:user-feedback-appendix}
In this section, we present feedback from two users on our theorem search engine.

\subsection{Feedback One}
We will first include a high-level description of the first feedback.

The user wanted to check for a specific reference, which we will call "Theorem 1", which the user knew was true. They thought it follows from another result, which we call ``result X,'' a bit stronger than Theorem 1. It did, after a small lemma; in other words, (result X + small lemma) implies what the user wanted. This shifted the user's attention from finding a reference for Theorem 1 to finding a reference for the small lemma. The user searched for the small lemma in our tool, and the first result was a reference for a generalization of the small lemma.

A more detailed description now follows. We advise the non-mathematician reader to skip the next paragraph.

The user wanted to check that a smooth variety over a separably closed field has a $k$-point. The user thought it follows from the \'etale local structure of smooth morphisms, which captured the correct idea but reduced the problem to showing that $k$-points are dense in $\mathbb{A}^n_k$. In essence, the \'etale local structure theorem translated the original problem of checking that a smooth variety over a separably closed field has a $k$-point, to checking that for $k$ separably closed, the set of $k$-points of $\mathbb{A}^n$ is dense. The user entered the query ``if $k$ is separably closed, the set of $k$-points of $\mathbb{A}^n$ is dense in $\mathbb{A}^n$.'' The first result was correct: Tag 056U in the Stacks Project.

\subsection{Feedback Two}
We now describe a second use case. As before, we first provide a high-level summary.

The user searched for a result known to experts but lacked a specific citable reference; we call this result ``Theorem~X.'' Many existing references point to similar results, but not exactly to Theorem~X. The user queried our tool and found a result, which we call ``Theorem~Y,'' whose citation list directly referenced Theorem~X. By using our tool, the user located the desired theorem with fewer queries than traditional search methods, which had previously led only to indirect references.

A more detailed description follows. We advise the non-mathematician reader to skip the next two paragraphs.

The user needed a structure theorem for abelian $p$-Lie algebras, as defined by N.\ Jacobson. The user conjectured that these algebras are classified analogously to finite abelian groups, provided the base field is algebraically closed. On MathOverflow, an answer directed the user to a book that states the desired result. The book explained that the result follows from the structure theory for modules over a PID, without providing a precise proof, and attributes the technique to Jacobson.

The book \textit{Infinite-Dimensional Lie Superalgebras} \cite{bahturin2011infinite} contains the result the user was looking for, labeled the 3.2 nil-radical Theorem, attributed to Seligman, 1967. However, the user skimmed through the hundred or so pages of Seligman's reference several times without finding where the desired result is proved. When the user queried our tool with ``over an algebraically closed field, any abelian $p$-Lie algebra splits as a torus and a sum of nilpotent cyclic Lie algebras,'' the first result was the desired statement: Theorem~2.7 in \cite{usefi2009isomorphisminvariantsrestrictedenveloping}.

\section{Claude Experiment}
\label{sec:claude_work}

We now report the experiment coupling Claude with our theorem search tool as a RAG system. This experiment tests whether retrieval-augmented generation over our theorem corpus can correct LLM reasoning failures on a research-level mathematical question. For discussion of this experiment in context, see Section~\ref{sec:claude-main}.

\begin{tcolorbox}[colback=gray!5, colframe=gray!60, title={\textbf{Query to Claude}}, fonttitle=\bfseries]
Consider the closure, in the KSBA moduli space, of the moduli space of pairs $(S,a B + bF)$ where
$S$ is an elliptic surface, $B$ is a section of the projection $S\to C$ which gives the elliptic surface structure to $S$, the fibers of $S\to C$ are smooth or nodal, $F$ are some fibers of $S\to C$, and $0<a,b< 1$.
The coefficients are chosen so that $(S,a B + bF)$ is KSBA-stable (so don't need to reprove it). I would like to know if there exist a surface pair $f:(X',aB'+bF')\to C'$ appearing on the boundary of the KSBA-compactification, not admitting a fibration over a nodal curve $C'$, with fibers of $f$ that are of dimension one.
\end{tcolorbox}

\noindent Without access to our tool, Claude answered \textbf{incorrectly}:

\begin{tcolorbox}[colback=red!3, colframe=red!50, title={\textbf{Claude's response without RAG (incorrect)}}, fonttitle=\bfseries]
\textbf{No.} Every surface pair appearing on the boundary of this KSBA compactification admits a fibration over a (possibly nodal) curve with one-dimensional fibers. The fibration structure cannot be lost in the limit.

\medskip
\textbf{Why the fibration persists.}
The essential point is that the elliptic fibration structure is rigidly encoded in the boundary divisor, and this encoding survives in the KSBA limit.

\medskip
\textbf{The fiber class remains nontrivial.} In the interior, $F$ is a fiber of $S \to C$, so $F^2 = 0$ and $(K_S + aB + bF) \cdot F > 0$ (since $K_S \cdot F = 0$, $B \cdot F = 1$, so this equals $a > 0$). In any flat limit, the limiting divisor $F'$ retains $F'^2 = 0$ and $(K_{X'} + aB' + bF') \cdot F' > 0$. So $F'$ is a nontrivial nef class with zero self-intersection, which defines a fibration $f: X' \to C'$.

\medskip
\textbf{Stable reduction preserves the fibered structure.} Given a one-parameter degeneration $(\mathcal{S}_t, \Delta_t) \to \mathcal{C}_t$, after finite base change and birational modification one extends the fibration to the central fiber. Each component of the limiting surface $X'$ either:
\begin{itemize}
\item maps surjectively onto a component of the nodal base curve $C'$ with one-dimensional fibers, or
\item maps to a point (a ``sprouted'' component).
\end{itemize}
But a component mapping to a point would be an entire surface sitting over a single point of $C'$, and the slc and ampleness conditions prevent this from occurring in the stable model: such a component would need $K+\Delta$ to be positive on it, yet its gluing locus with the rest of $X'$ is along a single fiber curve, and the combinatorics of the slc condition force it to be absorbed (contracted) when passing to the ample model of $K_{X'} + \Delta'$.

\medskip
\textbf{The section prevents collapse.} The section $B'$ (limit of the section $B$) meets each fiber in a point, ensuring the map $X' \to C'$ has generically one-dimensional fibers throughout. If any component lost the fibration structure, $B'$ restricted to that component would be a curve not meeting the fiber class transversally, contradicting the deformation-invariant intersection number $B \cdot F = 1$.

\medskip
\textbf{Summary.}
The divisor data $(aB + bF)$ is sufficiently rigid that all KSBA-stable limits retain a fibration $f: X' \to C'$ over a nodal curve $C'$, with all fibers of $f$ being one-dimensional (curves). The boundary pairs differ from interior pairs in that $C'$ may be nodal, $X'$ may be reducible/non-normal (slc), and fibers may be more degenerate---but the fibration itself always exists.
\end{tcolorbox}

\noindent When given access to our theorem database, Claude arrived at the \textbf{correct} answer:

\begin{tcolorbox}[colback=green!3, colframe=green!50!black, title={\textbf{Claude's response with RAG (correct)}}, fonttitle=\bfseries]
\textbf{Yes.} Such boundary pairs exist; they are pseudoelliptic surfaces.

\begin{itemize}
\item[\textbullet] By Corollary 1.3 of \cite{ascher2017logcanonicalmodelselliptic}, the log canonical model of
  an irreducible slc elliptic surface $(X\to C,\, S + F_{\mathcal{A}})$ with $K_X+S+F_{\mathcal{A}}$ big
  is either the relative log canonical model (an elliptic surface over $C$), or a
  \emph{pseudoelliptic surface} obtained by contracting the section $S$.

\item[\textbullet] By Corollary 6.10 of \cite{ascher2017logcanonicalmodelselliptic}, the section is contracted
  if and only if $(C, f_*F_{\mathcal{A}})$ is not a weighted stable pointed curve;
  concretely: $C\cong\mathbb{P}^1$ with $\sum a_i \le 2$, or $C$ has genus $1$ with all $a_i=0$.

\item[\textbullet] A pseudoelliptic surface admits no morphism to a curve with
  one-dimensional fibers: the section (which was isomorphic to the base) has been
  contracted to a point, so all former fibers pass through a single point and cannot
  be separated by a map to a curve.

\item[\textbullet] By Proposition 3.22 of \cite{ascher2018moduliweightedstableelliptic}, an explicit family is
  provided: for an elliptic surface over $\mathbb{P}^1$ with $\deg\mathscr{L}=2$ and
  $\mathcal{A}>0$, the log canonical model is a pseudoelliptic surface
  (section contracted), which is KSBA-stable and lies on the boundary.

\item[\textbullet] By Corollary 6.7 of \cite{inchiostro2018moduliweierstrassfibrationsmarked}, in a stable degeneration
  of elliptic surface pairs, the special fiber maps to a nodal curve $C'$, but
  pseudoelliptic components map to a \emph{point} of $C'$, so $f$ does not have
  all fibers of dimension one.

\item[\textbullet] Properness of the moduli space of slc fibered surface pairs
  (Corollary 5.11 of \cite{ascher2016modulifiberedsurfacepairs}) guarantees these pseudoelliptic limits
  lie in the closure.
\end{itemize}
\end{tcolorbox}

\clearpage

\section{Validation Set arXiv Categories}\label{val-set-arxiv-categories}

Table~\ref{tab:arxiv-tags} shows the distribution of arXiv categories across the 65 unique papers in our validation set. Out of 32 math tags on arXiv, the validation set covers 12.

\begin{table}[h]
\caption{Distribution of arXiv categories in the validation set. Each cell reports counts as \emph{primary\,/\,all}, where ``primary'' counts only papers whose first-listed arXiv category matches the row, and ``all'' also includes cross-listed papers.}
\label{tab:arxiv-tags}
\vskip 0.15in
\centering
\footnotesize
\setlength{\tabcolsep}{6pt}
\renewcommand{\arraystretch}{1.2}
\begin{tabular}{llcc}
\toprule
\textbf{arXiv Tag} & \textbf{Subject Area} & \textbf{Theorems} & \textbf{Papers} \\
\midrule
math.AG & Algebraic Geometry & 70\,/\,75 & 47\,/\,50 \\
math.CA & Classical Analysis and ODEs & 26\,/\,36 & 9\,/\,12 \\
math.AP & Analysis of PDEs & 17\,/\,23 & 4\,/\,7 \\
math.FA & Functional Analysis & --\,/\,12 & --\,/\,4 \\
math.MG & Metric Geometry & 7\,/\,9 & 2\,/\,3 \\
math.DG & Differential Geometry & --\,/\,9 & --\,/\,2 \\
math.NT & Number Theory & 2\,/\,7 & 1\,/\,4 \\
math.CT & Category Theory & --\,/\,4 & --\,/\,1 \\
math.CO & Combinatorics & 1\,/\,3 & 1\,/\,2 \\
math.RT & Representation Theory & --\,/\,3 & --\,/\,1 \\
nlin.SI & Exactly Solvable and Int. Sys. & 2\,/\,2 & 1\,/\,1 \\
math.CV & Complex Variables & --\,/\,1 & --\,/\,1 \\
math.SG & Symplectic Geometry & --\,/\,1 & --\,/\,1 \\
hep-th & High Energy Physics -- Theory & --\,/\,1 & --\,/\,1 \\
\bottomrule
\end{tabular}
\vskip -0.1in
\end{table}

\onecolumn
\newpage
\begin{longtable}{llcc}
\caption{Distribution of arXiv categories in the whole dataset. Each cell reports counts as \emph{primary\,/\,all}, where ``primary'' counts only papers whose first-listed arXiv category matches the row, and ``all'' also includes cross-listed papers.}
\label{tab:arxiv-tags-full}\\

\toprule
\textbf{arXiv Tag} & \textbf{Subject Area} & \textbf{Theorems} & \textbf{Papers} \\
\midrule
\endfirsthead

\multicolumn{4}{l}{\small Table \thetable\ (continued)}\\
\toprule
\textbf{arXiv Tag} & \textbf{Subject Area} & \textbf{Theorems} & \textbf{Papers} \\
\midrule
\endhead

\midrule
\multicolumn{4}{r}{\small Continued on next page}\\
\endfoot

\bottomrule
\endlastfoot

math.CO & Combinatorics & 727514\,/\,1107979 & 46929\,/\,68561 \\
math.AG & Algebraic Geometry & 761230\,/\,1097654 & 36350\,/\,51497 \\
math.AP & Analysis of PDEs & 791867\,/\,1018652 & 51176\,/\,64735 \\
math.PR & Probability & 606057\,/\,868961 & 38793\,/\,55971 \\
math.NT & Number Theory & 550466\,/\,752812 & 32235\,/\,42070 \\
math.DG & Differential Geometry & 499694\,/\,721447 & 28901\,/\,40763 \\
math.RT & Representation Theory & 353605\,/\,618985 & 15219\,/\,26364 \\
math.DS & Dynamical Systems & 374124\,/\,574125 & 21561\,/\,33646 \\
math.FA & Functional Analysis & 333709\,/\,544882 & 20193\,/\,31825 \\
math-ph & Mathematical Physics & 127036\,/\,533783 & 8604\,/\,32239 \\
math.GR & Group Theory & 303021\,/\,488750 & 14269\,/\,22702 \\
math.OC & Optimization and Control & 298096\,/\,470086 & 28546\,/\,43871 \\
math.GT & Geometric Topology & 277653\,/\,441500 & 15348\,/\,22650 \\
math.RA & Rings and Algebras & 202501\,/\,352448 & 10783\,/\,17650 \\
math.AT & Algebraic Topology & 200267\,/\,348076 & 8886\,/\,15601 \\
cs.LG & Machine Learning & 150507\,/\,323831 & 16346\,/\,33522 \\
math.QA & Quantum Algebra & 171100\,/\,322466 & 8258\,/\,15419 \\
math.CA & Classical Analysis and ODEs & 194189\,/\,309145 & 14237\,/\,21275 \\
math.NA & Numerical Analysis & 209943\,/\,285427 & 23450\,/\,30763 \\
math.LO & Logic & 206528\,/\,285351 & 9324\,/\,12952 \\
stat.ML & Machine Learning & 95327\,/\,281203 & 10123\,/\,28137 \\
math.CV & Complex Variables & 148649\,/\,278361 & 10203\,/\,16899 \\
math.OA & Operator Algebras & 170685\,/\,267922 & 8033\,/\,12677 \\
math.ST & Statistics Theory & 153124\,/\,260628 & 12979\,/\,21634 \\
math.AC & Commutative Algebra & 158613\,/\,253612 & 8762\,/\,13462 \\
math.MG & Metric Geometry & 100465\,/\,207104 & 6022\,/\,11402 \\
math.CT & Category Theory & 89204\,/\,202660 & 4024\,/\,8610 \\
math.SG & Symplectic Geometry & 110807\,/\,191221 & 4864\,/\,8954 \\
math.SP & Spectral Theory & 73394\,/\,162664 & 4666\,/\,9798 \\
math.KT & K-Theory and Homology & 57905\,/\,145433 & 2441\,/\,5961 \\
cs.DM & Discrete Mathematics & 35079\,/\,134746 & 2246\,/\,8829 \\
cs.CC & Computational Complexity & 65534\,/\,127305 & 4016\,/\,7808 \\
cs.DS & Data Structures and Algorithms & 53579\,/\,121709 & 3269\,/\,7629 \\
stat.ME & Methodology & 74267\,/\,117914 & 10991\,/\,15438 \\
cs.IT & Information Theory & 42758\,/\,109029 & 3422\,/\,8646 \\
hep-th & High Energy Physics – Theory & 17512\,/\,108924 & 1753\,/\,6809 \\
math.GN & General Topology & 60256\,/\,102259 & 3223\,/\,5212 \\
cs.AI & Artificial Intelligence & 24644\,/\,82598 & 3072\,/\,10378 \\
cs.LO & Logic in Computer Science & 30251\,/\,68125 & 1772\,/\,3945 \\
cs.CG & Computational Geometry & 33920\,/\,59488 & 2904\,/\,4755 \\
quant-ph & Quantum Physics & 25484\,/\,58911 & 1947\,/\,4332 \\
eess.SY & Systems and Control & 16997\,/\,52647 & 2549\,/\,7239 \\
gr-qc & General Relativity and Quantum Cosmology & 17771\,/\,43279 & 1125\,/\,2721 \\
cs.CR & Cryptography and Security & 16687\,/\,40367 & 2204\,/\,4419 \\
cs.GT & Computer Science and Game Theory & 17909\,/\,36310 & 1504\,/\,3096 \\
nlin.SI & Exactly Solvable and Integrable Systems & 8253\,/\,32897 & 744\,/\,2414 \\
math.GM & General Mathematics & 27194\,/\,29265 & 2224\,/\,2380 \\
stat.CO & Computation & 9804\,/\,28120 & 1515\,/\,3830 \\
econ.TH & Theoretical Economics & 19057\,/\,26467 & 1582\,/\,2167 \\
cond-mat.stat-mech & Statistical Mechanics & 2321\,/\,24482 & 287\,/\,1834 \\
q-fin.MF & Mathematical Finance & 14999\,/\,24160 & 1273\,/\,1960 \\
econ.EM & Econometrics & 14189\,/\,22632 & 1479\,/\,2234 \\
stat.AP & Applications & 5524\,/\,20294 & 1298\,/\,3778 \\
cs.FL & Formal Languages and Automata Theory & 6655\,/\,18854 & 439\,/\,1192 \\
eess.SP & Signal Processing & 10708\,/\,18805 & 2449\,/\,3485 \\
q-bio.PE & Populations and Evolution & 7196\,/\,17583 & 740\,/\,1743 \\
cs.DC & Distributed, Parallel, and Cluster Computing & 3689\,/\,16269 & 371\,/\,1669 \\
cs.SC & Symbolic Computation & 5435\,/\,14312 & 395\,/\,1113 \\
cs.SI & Social and Information Networks & 3772\,/\,13795 & 516\,/\,1550 \\
cs.CV & Computer Vision & 3247\,/\,12628 & 703\,/\,2169 \\
cs.MA & Multiagent Systems & 1522\,/\,12284 & 208\,/\,1386 \\
physics.flu-dyn & Fluid Dynamics & 1836\,/\,11450 & 393\,/\,1335 \\
math.HO & History and Overview & 5067\,/\,10829 & 427\,/\,815 \\
q-fin.CP & Computational Finance & 3495\,/\,8114 & 465\,/\,928 \\
nlin.CD & Chaotic Dynamics & 1715\,/\,7802 & 214\,/\,834 \\
cs.NE & Neural and Evolutionary Computing & 2010\,/\,7620 & 235\,/\,884 \\
q-fin.RM & Risk Management & 2840\,/\,7611 & 239\,/\,677 \\
q-fin.PR & Pricing of Securities & 3634\,/\,7537 & 357\,/\,710 \\
cs.RO & Robotics & 1972\,/\,7091 & 432\,/\,1192 \\
q-fin.PM & Portfolio Management & 3474\,/\,6878 & 296\,/\,626 \\
cond-mat.dis-nn & Disordered Systems and Neural Networks & 565\,/\,6180 & 67\,/\,504 \\
physics.comp-ph & Computational Physics & 1202\,/\,6038 & 313\,/\,1149 \\
econ.GN & General Economics & 3402\,/\,5953 & 446\,/\,695 \\
nlin.PS & Pattern Formation and Solitons & 1042\,/\,5864 & 121\,/\,518 \\
physics.soc-ph & Physics and Society & 1052\,/\,5639 & 211\,/\,819 \\
q-bio.QM & Quantitative Methods & 1417\,/\,5506 & 263\,/\,775 \\
cs.CL & Computation and Language & 1381\,/\,5300 & 277\,/\,767 \\
cs.NI & Networking and Internet Architecture & 1565\,/\,5000 & 222\,/\,664 \\
q-bio.MN & Molecular Networks & 1339\,/\,4549 & 134\,/\,408 \\
cs.CE & Computational Engineering, Finance, and Science & 1425\,/\,4295 & 280\,/\,731 \\
cond-mat.str-el & Strongly Correlated Electrons & 580\,/\,4123 & 55\,/\,275 \\
cs.DB & Databases & 1521\,/\,4043 & 144\,/\,415 \\
physics.data-an & Data Analysis, Statistics and Probability & 583\,/\,3747 & 111\,/\,541 \\
cs.PL & Programming Languages & 1120\,/\,3515 & 83\,/\,281 \\
physics.class-ph & Classical Physics & 1082\,/\,3288 & 178\,/\,459 \\
q-fin.ST & Statistical Finance & 1317\,/\,3221 & 197\,/\,421 \\
nlin.AO & Adaptation and Self-Organizing Systems & 558\,/\,3204 & 79\,/\,377 \\
q-bio.NC & Neurons and Cognition & 768\,/\,3054 & 102\,/\,378 \\
cs.PF & Performance & 973\,/\,3019 & 107\,/\,320 \\
cond-mat.mes-hall & Mesoscale and Nanoscale Physics & 231\,/\,2963 & 15\,/\,209 \\
cs.CY & Computers and Society & 405\,/\,2949 & 69\,/\,415 \\
cond-mat.mtrl-sci & Materials Science & 164\,/\,2924 & 29\,/\,271 \\
q-fin.TR & Trading and Market Microstructure & 1200\,/\,2919 & 107\,/\,269 \\
stat.OT & Other Statistics & 804\,/\,2794 & 111\,/\,339 \\
cs.IR & Information Retrieval & 564\,/\,2488 & 108\,/\,387 \\
physics.optics & Optics & 299\,/\,2033 & 40\,/\,211 \\
nlin.CG & Cellular Automata and Lattice Gases & 229\,/\,1860 & 22\,/\,136 \\
physics.plasm-ph & Plasma Physics & 265\,/\,1731 & 42\,/\,187 \\
q-fin.GN & General Finance & 391\,/\,1730 & 44\,/\,156 \\
cond-mat.soft & Soft Condensed Matter & 224\,/\,1594 & 39\,/\,168 \\
cond-mat.quant-gas & Quantum Gases & 49\,/\,1502 & 6\,/\,94 \\
cs.MS & Mathematical Software & 174\,/\,1487 & 55\,/\,281 \\
physics.chem-ph & Chemical Physics & 188\,/\,1353 & 32\,/\,164 \\
eess.IV & Image and Video Processing & 349\,/\,1350 & 74\,/\,282 \\
physics.bio-ph & Biological Physics & 96\,/\,1333 & 25\,/\,172 \\
cs.GR & Graphics & 254\,/\,1289 & 44\,/\,189 \\
physics.geo-ph & Geophysics & 79\,/\,1258 & 22\,/\,139 \\
physics.ao-ph & Atmospheric and Oceanic Physics & 113\,/\,1166 & 28\,/\,166 \\
physics.gen-ph & General Physics & 1010\,/\,1088 & 188\,/\,206 \\
hep-ph & High Energy Physics – Phenomenology & 31\,/\,957 & 6\,/\,93 \\
cs.ET & Emerging Technologies & 194\,/\,876 & 27\,/\,102 \\
hep-lat & High Energy Physics – Lattice & 42\,/\,831 & 9\,/\,72 \\
q-bio.GN & Genomics & 179\,/\,811 & 27\,/\,125 \\
cs.HC & Human-Computer Interaction & 128\,/\,760 & 20\,/\,121 \\
q-bio.BM & Biomolecules & 176\,/\,701 & 24\,/\,92 \\
physics.med-ph & Medical Physics & 117\,/\,619 & 24\,/\,94 \\
astro-ph.IM & Instrumentation and Methods for Astrophysics & 64\,/\,586 & 15\,/\,83 \\
q-bio.CB & Cell Behavior & 99\,/\,555 & 17\,/\,83 \\
astro-ph.CO & Cosmology and Nongalactic Astrophysics & 82\,/\,554 & 4\,/\,54 \\
q-bio.TO & Tissues and Organs & 64\,/\,482 & 12\,/\,73 \\
cond-mat.other & Other Condensed Matter & 19\,/\,441 & 2\,/\,39 \\
cond-mat.supr-con & Superconductivity & 22\,/\,434 & 3\,/\,26 \\
physics.app-ph & Applied Physics & 20\,/\,431 & 6\,/\,64 \\
cs.SE & Software Engineering & 147\,/\,406 & 30\,/\,80 \\
cond-mat & Condensed Matter & 83\,/\,383 & 9\,/\,42 \\
physics.hist-ph & History and Philosophy of Physics & 116\,/\,380 & 23\,/\,54 \\
cs.SD & Sound & 95\,/\,373 & 28\,/\,74 \\
astro-ph.EP & Earth and Planetary Astrophysics & 121\,/\,366 & 23\,/\,55 \\
physics.atom-ph & Atomic Physics & 4\,/\,316 & 1\,/\,24 \\
cs.AR & Hardware Architecture & 37\,/\,314 & 6\,/\,36 \\
eess.AS & Audio and Speech Processing & 7\,/\,299 & 3\,/\,66 \\
astro-ph & Astrophysics & 38\,/\,272 & 4\,/\,30 \\
q-bio.SC & Subcellular Processes & 61\,/\,257 & 10\,/\,33 \\
cs.OH & Other Computer Science & 22\,/\,246 & 3\,/\,22 \\
nucl-th & Nuclear Theory & 2\,/\,232 & 2\,/\,24 \\
cs.MM & Multimedia & 18\,/\,190 & 7\,/\,46 \\
astro-ph.SR & Solar and Stellar Astrophysics & 33\,/\,150 & 7\,/\,26 \\
cs.DL & Digital Libraries & 29\,/\,104 & 11\,/\,21 \\
physics.acc-ph & Accelerator Physics & 12\,/\,104 & 3\,/\,16 \\
physics.space-ph & Space Physics & 6\,/\,82 & 2\,/\,12 \\
hep-ex & High Energy Physics – Experiment & 1\,/\,39 & 1\,/\,10 \\
q-bio.OT & Other Quantitative Biology & 4\,/\,36 & 1\,/\,8 \\
cs.OS & Operating Systems & 16\,/\,25 & 5\,/\,7 \\
physics.pop-ph & Popular Physics & 3\,/\,17 & 2\,/\,5 \\
physics.ins-det & Instrumentation and Detectors & 1\,/\,15 & 1\,/\,4 \\

\end{longtable}

\section{More Experiments}
\label{more-experiments}

We ran our experiments with a larger set of embedders, including Multilingual-E5-Large-Instruct \cite{wang2024multilingual}, zbMath-Bert \cite{zbmath}, and KaLM-Embedding-V2.5 \cite{zhao2025kalmembeddingv2, hu2025kalmembedding}.

\begin{table}[ht!]
\caption{Extended results comparing context window size in retrieval performance. Embedded without task instructions.}
\label{context-window-table-full}
\vskip 0.15in
\begin{center}
\begin{small}
\begin{sc}
\resizebox{\columnwidth}{!}{
\begin{tabular}{lcccc}
\toprule
Model & Hit@1 & Hit@10 & Hit@20 & MRR@20 \\
\midrule
\multicolumn{5}{c}{Body Only}\\
\midrule
zbMath-Bert 0.3B & 0.053 & 0.158 & 0.237 & 0.086 \\
Gemma 0.3B & 0.184 & 0.474 & 0.539 & 0.290 \\
Qwen3 0.6B & 0.224 & 0.461 & 0.513 & 0.299 \\
Qwen3 8B & 0.224 & 0.553 & 0.632 & 0.313 \\
KaLM-Embedding-V2.5 & 0.131 & 0.355 & 0.434 & 0.203 \\
Multilingual-E5-Large-Instruct & 0.171 & 0.395 & 0.500 & 0.237 \\
\midrule
\multicolumn{5}{c}{w/ Abstract}\\
\midrule
zbMath-Bert 0.3B & 0.092 & 0.184 & 0.237 & 0.113 \\
Gemma 0.3B & 0.303 & 0.539 & 0.566 & 0.376 \\
Qwen3 0.6B & 0.224 & 0.487 & 0.526 & 0.297 \\
Qwen3 8B & 0.250 & 0.553 & 0.618 & 0.332 \\
KaLM-Embedding-V2.5 & 0.197 & 0.474 & 0.513 & 0.272 \\
Multilingual-E5-Large-Instruct & 0.197 & 0.421 & 0.553 & 0.283 \\
\midrule
\multicolumn{5}{c}{w/ First Section}\\
\midrule
zbMath-Bert 0.3B & 0.092 & 0.184 & 0.237 & 0.122 \\
Gemma 0.3B & 0.237 & 0.500 & 0.566 & 0.321 \\
Qwen3 0.6B & 0.263 & 0.526 & 0.645 & 0.352 \\
Qwen3 8B & 0.211 & 0.605 & 0.645 & 0.343 \\
KaLM-Embedding-V2.5 & 0.184 & 0.447 & 0.474 & 0.272 \\
Multilingual-E5-Large-Instruct & 0.224 & 0.474 & 0.539 & 0.304 \\
\bottomrule
\end{tabular}
}
\end{sc}
\end{small}
\end{center}
\vskip -0.1in
\end{table}

\begin{table}[ht!]
\caption{Extended results comparing LLM slogans in retrieval performance. Embedded without task instructions. Body+Abstract.}
\label{slogan-llm-table-full}
\vskip 0.15in
\begin{center}
\begin{small}
\begin{sc}
\resizebox{\columnwidth}{!}{
\begin{tabular}{lcccc}
\toprule
Model & Hit@1 & Hit@10 & Hit@20 & MRR@20 \\
\midrule
\multicolumn{5}{c}{Deepseek V3.1}\\
\midrule
zbMath-Bert 0.3B & 0.092 & 0.184 & 0.237 & 0.113 \\
Gemma 0.3B & 0.303 & 0.539 & 0.566 & 0.376 \\
Qwen3 0.6B & 0.224 & 0.487 & 0.526 & 0.297 \\
Qwen3 8B & 0.250 & 0.553 & 0.618 & 0.332 \\
KaLM-Embedding-V2.5 & 0.197 & 0.474 & 0.513 & 0.272 \\
Multilingual-E5-Large-Instruct & 0.197 & 0.421 & 0.553 & 0.283 \\
\midrule
\multicolumn{5}{c}{Deepseek R1}\\
\midrule
zbMath-Bert 0.3B & 0.000 & 0.079 & 0.105 & 0.014 \\
Gemma 0.3B & 0.158 & 0.421 & 0.474 & 0.228 \\
Qwen3 0.6B & 0.132 & 0.316 & 0.408 & 0.199 \\
Qwen3 8B & 0.118 & 0.355 & 0.461 & 0.195 \\
KaLM-Embedding-V2.5 & 0.066 & 0.289 & 0.461 & 0.140 \\
Multilingual-E5-Large-Instruct & 0.066 & 0.250 & 0.355 & 0.111 \\
\midrule
\multicolumn{5}{c}{Gemini 3 Pro}\\
\midrule
zbMath-Bert 0.3B & 0.053 & 0.171 & 0.237 & 0.092 \\
Gemma 0.3B & 0.250 & 0.553 & 0.632 & 0.348 \\
Qwen3 0.6B & 0.211 & 0.487 & 0.539 & 0.290 \\
Qwen3 8B & 0.237 & 0.566 & 0.684 & 0.348 \\
KaLM-Embedding-V2.5 & 0.197 & 0.434 & 0.553 & 0.265 \\
Multilingual-E5-Large-Instruct & 0.197 & 0.513 & 0.579 & 0.297 \\
\midrule
\multicolumn{5}{c}{Claude Opus 4.5}\\
\midrule
zbMath-Bert 0.3B & 0.053 & 0.184 & 0.211 & 0.087 \\
Gemma 0.3B & 0.303 & 0.553 & 0.579 & 0.387 \\
Qwen3 0.6B & 0.224 & 0.539 & 0.605 & 0.325 \\
Qwen3 8B & 0.263 & 0.632 & 0.737 & 0.394 \\
KaLM-Embedding-V2.5 & 0.197 & 0.421 & 0.526 & 0.271 \\
Multilingual-E5-Large-Instruct & 0.224 & 0.526 & 0.592 & 0.322 \\
\bottomrule
\end{tabular}
}
\end{sc}
\end{small}
\end{center}
\vskip -0.1in
\end{table}

\clearpage
\FloatBarrier

\section{Prompts}
In this section we list the prompts used during slogan generation and retrieval, shown in Tables~\cref{doc-prep-table-describe}, \cref{slogan-prompts}, and~\cref{retrieval-prompts}.

\begin{table}[!ht]
\caption{Task instructions for embedders}
\label{doc-prep-table-describe}
\vskip 0.15in
\begin{center}
\footnotesize
\begin{sc}
\begin{tabular}{l p{0.65\columnwidth}}
\toprule
Side & Prompt \\
\midrule
Theorem &
\texttt{Instruct: Represent the given math statement for retrieving related statement by natural language query.\textbackslash nQuery:} \\
User Query &
\texttt{Instruct: Given a math search query, retrieve theorems mathematically equivalent to the query.\textbackslash nQuery:} \\
\bottomrule
\end{tabular}
\end{sc}
\end{center}
\vskip -0.1in
\end{table}

\begin{table}[!ht]
\caption{Slogan prompts}
\label{slogan-prompts}
\vskip 0.15in
\centering
\footnotesize
\setlength{\tabcolsep}{6pt}
\renewcommand{\arraystretch}{1.2}

\begin{tabular}{p{0.18\textwidth} p{0.78\textwidth}}
\toprule
\textbf{Context} & \textbf{Prompt} \\
\midrule

Body Only &
\ttfamily
You generate summaries of math theorems based on theorem\_body. Summaries are accurate and at most four sentences. Summaries are plain ASCII sentences with no Unicode. Describe the result without referencing it as 'this theorem' or similar. Avoid LaTeX and mathematical symbols; use words instead. Output only the final summary sentences, with no reasoning, explanations, or commentary. Do not restate the prompt, input fields, or instructions. Do not include proof steps, motivation, or background discussion.
\normalfont
\\

Body+Abstract &
\ttfamily
You generate summaries of math theorems based on theorem\_body. You also consider paper\_summary in your summaries. Summaries are accurate and at most four sentences. Summaries are plain ASCII sentences with no Unicode. Describe the result without referencing it as 'this theorem' or similar. Avoid LaTeX and mathematical symbols; use words instead. Output only the final summary sentences, with no reasoning, explanations, or commentary. Do not restate the prompt, input fields, or instructions. Do not include proof steps, motivation, or background discussion.
\normalfont
\\

Body+Introduction &
\ttfamily
You generate summaries of math theorems based on theorem\_body. You also consider paper\_summary and the first section of the paper in your summaries. Summaries are accurate and at most four sentences. Summaries are plain ASCII sentences with no Unicode. Describe the result without referencing it as 'this theorem' or similar. Avoid LaTeX and mathematical symbols; use words instead. Output only the final summary sentences, with no reasoning, explanations, or commentary. Do not restate the prompt, input fields, or instructions. Do not include proof steps, motivation, or background discussion.
\normalfont
\\

\bottomrule
\end{tabular}
\vskip -0.1in
\end{table}

\begin{table*}[]
\caption{Retrieval prompts for baselines.}
\label{retrieval-prompts}
\vskip 0.15in
\centering
\footnotesize
\setlength{\tabcolsep}{6pt}
\renewcommand{\arraystretch}{1.2}
\begin{sc}
\begin{tabular}{p{0.18\textwidth} p{0.78\textwidth}}
\toprule
Context & Prompt \\
\midrule

GPT-5.2 &
{\ttfamily\raggedright
\{\\
\ \ ``role'': ``system'',\\
\ \ ``content'': (\\
\ \ \ \ ``You are an assistant that MUST answer immediately.\textbackslash n''\\
\ \ \ \ ``Do NOT ask the user questions or request permission.\textbackslash n''\\
\ \ \ \ ``Use web search as needed, including opening the latest arXiv PDF(s) to verify statement numbering.\textbackslash n''\\
\ \ \ \ ``If exact numbering cannot be verified for some item, still include it but mark number as 'UNVERIFIED'.\textbackslash n''\\
\ \ \ \ ``Return ONLY the final list—no preamble.''\\
\ \ ),\\
\},\\
\{\\
\ \ ``role'': ``user'',\\
\ \ ``content'': (\\
\ \ \ \ ``Return a list of the top 20 most relevant math statements to the query below.\textbackslash n''\\
\ \ \ \ ``Constraints:\textbackslash n''\\
\ \ \ \ ``- Statements must be from arXiv papers.\textbackslash n''\\
\ \ \ \ ``- For each item include: (1) arXiv id, (2) version used (e.g. v3), ''\\
\ \ \ \ ``(3) statement type+number exactly as in that arXiv version (e.g. Theorem 1.2 / Lemma 3.4), ''\\
\ \ \ \ ``(4) section name/number, (5) statement title/short descriptor, (6) 1–2 sentence relevance note.\textbackslash n''\\
\ \ \ \ ``- Use the most recent arXiv version available.\textbackslash n''\\
\ \ \ \ ``Query:\textbackslash n''\\
\ \ \ \ f``\{row.query\}''\\
\ \ )\\
\}\\
}
\\

Gemini 3 Pro &
{\ttfamily\raggedright
SYSTEM\_INSTRUCTION = (\\
\ \ \ \ ``You are an assistant that MUST answer immediately.\textbackslash n''\\
\ \ \ \ ``Do NOT ask the user questions or request permission.\textbackslash n''\\
\ \ \ \ ``Use Google Search grounding as needed, including opening the latest arXiv PDF(s) to verify statement numbering.\textbackslash n''\\
\ \ \ \ ``If exact numbering cannot be verified for some item, still include it but mark number as 'UNVERIFIED'.\textbackslash n''\\
\ \ \ \ ``Return ONLY the final list—no preamble.''\\
)\\[0.5em]
user\_prompt = (\\
\ \ \ \ ``Return a list of the top 20 most relevant math statements to the query below.\textbackslash n''\\
\ \ \ \ ``Constraints:\textbackslash n''\\
\ \ \ \ ``- Statements must be from arXiv papers.\textbackslash n''\\
\ \ \ \ ``- For each item include: (1) arXiv id, (2) version used (e.g. v3), ''\\
\ \ \ \ ``(3) statement type+number exactly as in that arXiv version (e.g. Theorem 1.2 / Lemma 3.4), ''\\
\ \ \ \ ``(4) section name/number, (5) statement title/short descriptor, (6) 1–2 sentence relevance note.\textbackslash n''\\
\ \ \ \ ``- Use the most recent arXiv version available.\textbackslash n''\\
\ \ \ \ ``Query:\textbackslash n''\\
\ \ \ \ f``\{row.query\}''\\
)\\
}
\\

\bottomrule
\end{tabular}
\end{sc}
\vskip -0.1in
\end{table*}


\end{document}